\documentclass[aps,prx,superscriptaddress,amsmath,amssymb,twocolumn,showpacs,floatfix,reprint]{revtex4-2}
\usepackage[colorlinks=true, urlcolor=blue, linkcolor=blue, citecolor=blue, pdftex]{hyperref}
\usepackage[capitalise]{cleveref}
\usepackage{xr} 
\usepackage[dvipsnames]{xcolor}
\usepackage[utf8]{inputenc}
\usepackage{braket}
\usepackage{graphicx}

\definecolor{C0}{HTML}{1f77b4}
\definecolor{C1}{HTML}{ff7f0e}
\definecolor{C2}{HTML}{2ca02c}
\definecolor{C3}{HTML}{d62728}
\definecolor{C4}{HTML}{9467bd}
\definecolor{C5}{HTML}{8c564b}

\let\b\boldsymbol

\begin{document}

\title{Quantum Spin Glass in the Two-Dimensional Disordered Heisenberg Model via Foundation Neural-Network Quantum States}

\date{\today}

\begin{abstract}
We investigate the two-dimensional frustrated quantum Heisenberg model with bond disorder on nearest-neighbor couplings using the recently introduced Foundation Neural-Network Quantum States framework, which enables accurate and efficient computation of disorder-averaged observables with a single variational optimization. Simulations on large lattices reveal an extended region of the phase diagram where conventional magnetic long-range order vanishes in the thermodynamic limit, while the Edwards-Anderson order parameter remains finite, signaling the emergence of a quantum spin-glass phase. These findings, supported by a semiclassical analysis based on a large-spin expansion, provide compelling evidence that the spin glass-order is stable against quantum fluctuations, unlike the classical case where it disappears at any finite temperature.
\end{abstract}

\author{Luciano Loris Viteritti}
\email{luciano.viteritti@epfl.ch}
\affiliation{Institute of Physics, \'{E}cole Polytechnique F\'{e}d\'{e}rale de Lausanne (EPFL), CH-1015 Lausanne, Switzerland}

\author{Riccardo Rende}
\affiliation{SISSA, via Bonomea 265, 34136, Trieste, Italy}

\author{Giacomo Bracci Testasecca}
\email{gbraccit@sissa.it}
\affiliation{SISSA, via Bonomea 265, 34136, Trieste, Italy}
\affiliation{INFN, Sezione di Trieste, Via Valerio 2, 34127 Trieste, Italy}

\author{\\ Jacopo Niedda}
\email{jniedda@ictp.it}
\affiliation{The Abdus Salam ICTP, Strada Costiera 11, 34151 Trieste, Italy}

\author{Roderich Moessner}
\affiliation{Max Planck Institute for the Physics of Complex Systems, Nöthnitzer Str. 38, 01187 Dresden, Germany}

\author{Giuseppe Carleo}
\affiliation{Institute of Physics, \'{E}cole Polytechnique F\'{e}d\'{e}rale de Lausanne (EPFL), CH-1015 Lausanne, Switzerland}

\author{Antonello Scardicchio}
\email{ascardic@ictp.it}
\affiliation{The Abdus Salam ICTP, Strada Costiera 11, 34151 Trieste, Italy}
\affiliation{INFN, Sezione di Trieste, Via Valerio 2, 34127 Trieste, Italy}

\maketitle
\section{Introduction} 
Spin glasses (SG) have long been a fascinating topic of research in both theoretical and experimental physics~\cite{Mezard87,mydosh1993spin}. Introduced in the 1970s~\cite{edwards1975theory} in the context of magnetic alloys with quenched impurities, SG exhibit a low-temperature behavior that is markedly different from ferromagnetism, antiferromagnetism, or other, more readily recognizable forms of order. What appears as a random freezing of spins in these materials actually conceals a subtle form of long-range order, revealed by spatio-temporal correlations among the local degrees of freedom. The spin-glass phase has been rigorously characterized within mean-field theory~\cite{Parisi79a,Parisi80a,Parisi80b,Bray1980,Mezard84}, which predicts the existence of a multitude of equilibrium states, not related by symmetry. This richness gives rise to a variety of intriguing out-of-equilibrium phenomena, including weak ergodicity breaking and aging~\cite{Cugliandolo1997}, and makes SG a paradigmatic example of complex systems with far-reaching implications across many areas of research \cite{parisi2023nobel}.

Of particular interest is the regime in which, upon lowering the temperature of a many-body system with quenched disorder, quantum effects become relevant. In this regime, one speaks of quantum spin glasses (QSG)~\cite{Cugliandolo23}. Even if quantum fluctuations might, in principle, destroy order, experiments strongly suggest that spin-glass order persists at the quantum level, albeit with reduced transition temperatures and shorter characteristic relaxation times~\cite{Wu1991,Wu1993}. Many of the hallmark features of classical SG, including slow relaxation and aging~\cite{Cugliandolo99,Kennet01a,Kennet01b}, survive in the quantum regime, where they coexist with genuinely quantum phenomena. QSG are a natural platform to investigate the interplay of disorder and frustration in quantum systems. In particular, they may exhibit parameter chaos~\cite{Baity-Jesi2021}, namely the sudden change of the ground-state configuration induced by infinitesimal variations of a control parameter and, possibly, a breakdown of ergodicity, \emph{e.g.}~signaled by a violation of the eigenstate thermalization hypothesis~\cite{srednicki1994chaos,deutsch2018eigenstate,Baldwin2016,Mossi2017,balducci2025}. In the extreme disorder limit, they may also prevent equilibration altogether~\cite{sierant2025many}. Notably, QSG dynamics can also serve as a model for the behavior of quantum algorithms~\cite{Santoro2002,laumann2012quantum,laumann2015quantum,Martin-Mayor2015, Katzgraber2015} and, therefore, are relevant to a theory of quantum complexity. More broadly, the interplay of disorder and quantum fluctuations has been proposed as a mechanism capable of suppressing quasiparticle excitations and stabilizing non-Fermi-liquid
phases down to zero temperature~\cite{maldacena2016remarks,sachdev1993gapless}. 

The theoretical description of SG, both classical and quantum, presents numerous challenges that are resilient to both analytic and numerical treatments. Averaging the observables over the disorder, {\it i.e.}, computing {\it quenched averages}, analytically requires either the introduction of replicas or auxiliary fermionic fields. While the latter technique seems more powerful when one aims at describing localization physics in disordered systems~\cite{Efetov96}, the former has proven to be useful in devising the mean-field description of the spin glass phase in terms of Replica Symmetry Breaking (RSB)~\cite{Mezard87}. However, neither method allows one to solve for finite-dimensional SG, a situation that has led to controversies regarding the applicability of mean-field theory in low dimensions or the values of the upper and lower critical dimensions of the spin glass transition \cite{Moore2005,DeDominicis2006,Bray2011,RuizLorenzo2020,Angelini2022}. 

To address the question of whether spin-glass order survives in finite-dimensional quantum systems, we investigate the ground-state properties of the two-dimensional quantum Heisenberg model on a square lattice with binary bond disorder in nearest-neighbor couplings. Besides its intrinsic relevance to the theory of QSG, the study of this system is experimentally motivated by the physics of the copper-oxygen planes in lightly doped insulating high-$T_c$ superconductors. According to Refs.~\cite{aharony1988,aharony1988b,frachet2020}, localized vacancies on oxygen sites induce effective ferromagnetic couplings between neighboring copper spins, generating magnetic frustration and potentially leading to spin-glass behavior.

In general, studying frustrated disordered quantum magnets in two dimensions is notoriously challenging. Quantum Monte Carlo (QMC) algorithms are often hindered by the sign problem, while exact diagonalization is restricted to small system sizes~\cite{oitmaa2001,arrachea2001,rodriguez1995,uematsu2018,wu2024}. Density Matrix Renormalization Group (DMRG) and related tensor-network techniques have proven highly accurate even in two-dimensional disordered systems~\cite{ren2023}, but they face significant limitations compared to their one-dimensional counterparts. These include the need for high-rank tensor structures or, alternatively, restrictions to quasi-one-dimensional geometries using low-rank tensors arranged along a snaked path~\cite{stoudenmire2012}. Moreover, such approaches often struggle to efficiently implement fully periodic boundary conditions, which are important for mitigating finite-size effects. Regardless of the computational method employed, one of the main challenges in studying disordered systems lies in the large number of realizations required to precisely compute disorder-averaged observables.

As a consequence, the numerical study of quantum disordered magnets has been restricted to a few special cases. A prominent example is the Ising spin glass in a transverse field, where, since the work of Rieger and Young \cite{Rieger1994}, extensive sign-problem-free QMC simulations have established the persistence of spin-glass order up to substantial values of the transverse field~\cite{Rieger1994,choi2023,kramer2024,Bernaschi2024}. Establishing the stability of spin-glass order in the context of disordered Heisenberg models, which are generally affected by the sign problem, would provide nontrivial evidence that the spin-glass phase survives fully $SU(2)$-symmetric quantum fluctuations. Beyond the transverse-field Ising case, only a limited number of disordered quantum magnets have been studied numerically. These include two-dimensional bipartite antiferromagnets with weak bond disorder~\cite{sandvik1994} and limited random site or bond dilution~\cite{sandvik2002}, or the  one-dimensional disordered Heisenberg model~\cite{fava2024,sibei2025}. Interestingly, the latter system has recently become the subject of an active debate. While spin-wave and DMRG studies have provided evidence in favor of spin-glass order~\cite{fava2024}, recent ultra-low-temperature sign-free QMC simulations have instead suggested an unconventional critical state with power-law correlations~\cite{sibei2025}. At the same time, mean-field theory provides evidence that quantum fluctuations do not necessarily destroy spin-glass order. A quantum spin-glass phase in the fully connected Heisenberg model was already found by Bray and Moore~\cite{Bray1980} for every value $S$ of the spin magnitude under assumption of unbroken replica symmetry (RS). The full RSB solution for $S=\tfrac{1}{2}$ has instead only recently been found in Ref.~\cite{kavokine2024exact}. Whether spin-glass order survives in these models in finite dimensions remains an open question, which we aim to address in this work.

\begin{figure}
    \centering
    \includegraphics[width=\linewidth]{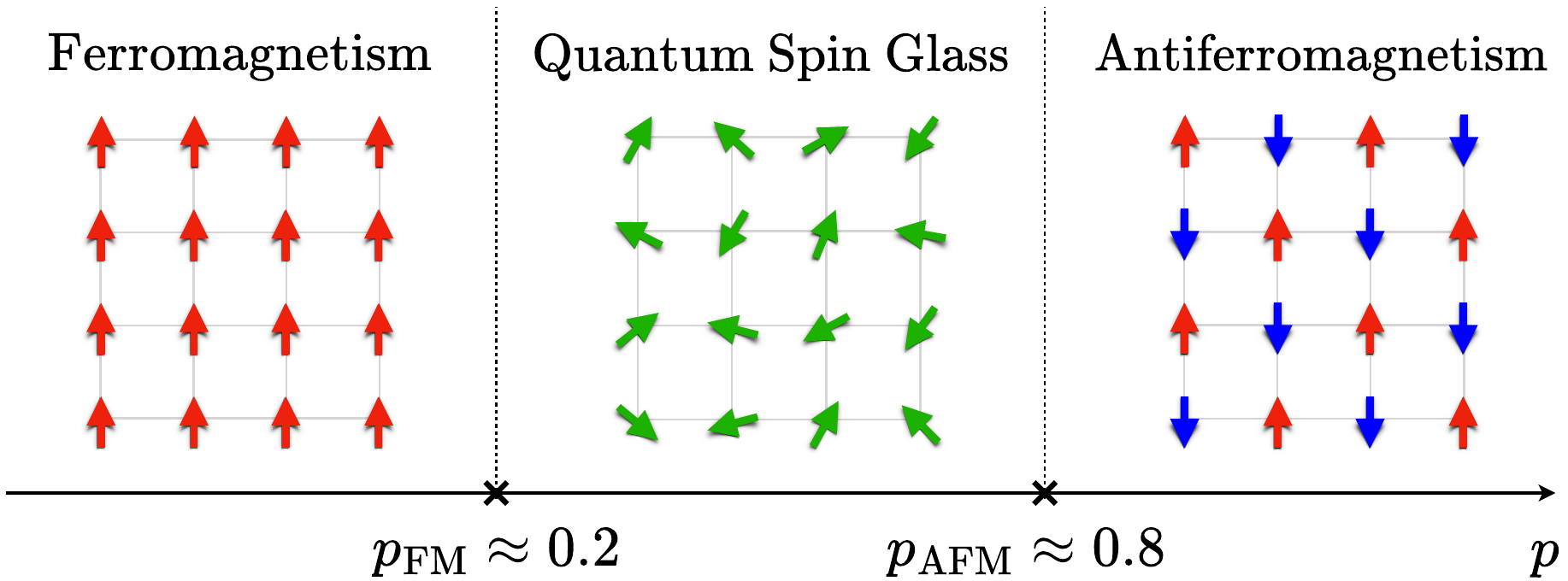}
    \caption{Ground-state phase diagram of the disordered Heisenberg model [see Eq.~\eqref{eq:ham_model}] in the thermodynamic limit, as a function of the probability $p \in [0,1]$ [see Eq.~\eqref{eq:prob_couplings}]. Three distinct phases are identified: a ferromagnetic phase for ${p \le p_{\mathrm{FM}} \approx 0.2}$, an antiferromagnetic phase for ${p \ge p_{\mathrm{AFM}} \approx 0.8}$, and an intermediate quantum spin glass phase characterized by the absence of magnetic order and a finite Edwards-Anderson order parameter $Q$ (see \emph{Order parameters} for details). The phase diagram is obtained using the variational approach based on Foundation Neural-Network Quantum States (refer to \emph{Methods}).}
    \label{fig:phase_diagram}
\end{figure}

To tackle this problem, we employ the Foundation Neural-Network Quantum States (FNQS) variational framework introduced in Ref.~\cite{rende2025foundation}. This approach allows us to circumvent the sign problem, retain fully periodic boundary conditions, and efficiently perform disorder averaging on lattice sizes that are inaccessible to exact diagonalization. The central result of this study is the identification of an intermediate phase, characterized as a QSG state, between a ferromagnetic and an antiferromagnetic phase by tuning the disorder strength. A schematic phase diagram of the model summarizing these findings is presented in Fig.~\ref{fig:phase_diagram}. To further support our conclusions, we perform a semiclassical analysis based on a large-spin expansion at the leading order in $1/S$. This analysis shows that the classical spin-glass order found at $T=0$, which is known to be unstable for any $T>0$~\cite{Fernandez1977,Kawamura03_2D}, is instead stable against quantum fluctuations. In particular, we find that the leading $1/S$ correction to the spin-glass order parameter yields a value compatible with the quantum value obtained through the FNQS approach in the thermodynamic limit, supporting the existence of a stable QSG phase at $S = \tfrac{1}{2}$, as indicated by our fully quantum numerical simulations. 

\section{The disordered Heisenberg model} 
The two-dimensional Heisenberg model with binary random couplings on an $L \times L$ square lattice is defined by:
\begin{equation}\label{eq:ham_model}
    \hat{H} = \sum_{\langle i,j\rangle} J_{ij} \hat{\boldsymbol{S}}_i \cdot \hat{\boldsymbol{S}}_j \ ,
\end{equation}
where $\hat{\boldsymbol{S}}_{i}=(\hat{{S}}^x_{i},\hat{{S}}^y_{i},\hat{{S}}^z_{i})$ is the $S=\tfrac{1}{2}$ spin operator at site $i$, and the sum runs only over nearest-neighbor sites, assuming periodic boundary conditions. The exchange couplings $J_{ij}$ are randomly distributed according to the probability distribution:
\begin{equation}\label{eq:prob_couplings}
    P(J_{ij}) = (1-p) \delta(J_{ij} + 1) + p \delta (J_{ij} - 1) \ ,
\end{equation}
being $p \in [0,1]$ the probability of having an antiferromagnetic bond between site $i$ and $j$. The ground state is well understood in two limiting cases: at $p = 1$, where the model reduces to the antiferromagnetic Heisenberg model, exhibiting long-range Néel order~\cite{calandra1998,sandvik1997}, and at $p = 0$, where ferromagnetic order dominates. However, for a generic value of $0 < p < 1$, this model is notably challenging to simulate with standard techniques, and in the highly frustrated regime $p \approx 1/2$, the nature of the ground state remains elusive.

\begin{figure*}
    \begin{center}
    \centerline{\includegraphics[width=2.\columnwidth]{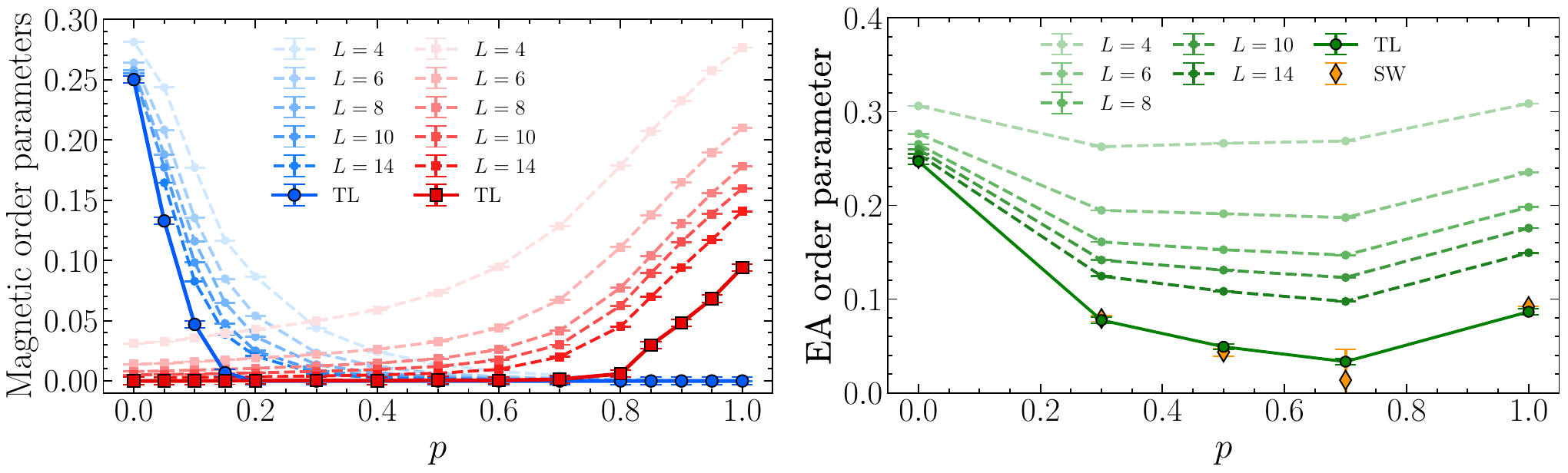}}
        \caption{\textbf{Left panel:} Ferromagnetic order parameter $\mathcal{M}^2_{\text{Ferro}}$ (blue circles, dashed lines) and N\'eel antiferromagnetic order parameter $\mathcal{M}^2_{\text{N\'eel}}$ (red squares, dashed lines) are shown as a function of the probability $p$ for system sizes ranging from $L=4$ to $L=14$. Extrapolated values in the thermodynamic limit (TL), obtained via finite-size scaling in $1/L$ (see \cref{sec:extrapolations} of the \textit{Methods} for details), are indicated by solid lines connecting the corresponding symbols. 
        \textbf{Right panel:} Edwards-Anderson order parameter $Q$ (green circles, dashed lines) as a function of $p$ for the same system sizes and averaging procedure. Extrapolated TL values are shown as green circles connected by a solid line.
        Results in both panels are averaged over $\mathcal{R} = 600$ disorder realizations. The error on the extrapolated values in the thermodynamic limit are estimated via a resampling technique with Gaussian noise. The semiclassical value of the Edwards-Anderson parameter obtained in the TL of the non-interacting spin-wave theory is shown for comparison (orange rhombi), in great agreement with the FNQS values (see \emph{Semiclassical analysis} for the details).}
        \label{fig:order_params}
    \end{center}
\end{figure*}

Previous studies~\cite{oitmaa2001,arrachea2001,rodriguez1995} have suggested the possibility of a QSG phase; however, these findings were limited by exact diagonalization results obtained on small $4 \times 4$ clusters and are far from compelling. In Ref.~\cite{sandvik1994}, square-lattice antiferromagnets were studied at low temperatures on clusters of up to $10 \times 10$ sites. However, only up to $10\%$ ferromagnetic bonds were considered, limiting access to the spin-glass regime of primary interest in the present work (which we show appears when the fraction of ferromagnetic bonds exceeds $20\%$). At the mean-field level, a low temperature SG phase has been recently found on the fully-connected graph with Gaussian distributed couplings in Ref.~\cite{kavokine2024exact}, through a sophisticated dynamical mean-field method, which includes Replica Symmetry Breaking and a continuous time Quantum Monte Carlo. However, as usual, the applicability of mean-field theory to low dimensions can be questioned.\\

\section{Results} 
We adopt the methodology introduced in Ref.~\cite{rende2025foundation}, in which a variational quantum state
$|\psi_{\theta}(\boldsymbol{J})\rangle$, parametrized by a neural network and referred to as a Foundation Neural-Network Quantum State (FNQS), is optimized by minimizing the disorder-averaged energy:
\begin{equation}\label{eq:loss}
    \mathcal{L}(\theta)
    =
    \mathbb{E}
    \left[
    \frac{
    \braket{\psi_{\theta}(\boldsymbol{J})|\hat{H}_{\boldsymbol{J}}|\psi_{\theta}(\boldsymbol{J})}
    }{
    \braket{\psi_{\theta}(\boldsymbol{J})|\psi_{\theta}(\boldsymbol{J})}
    }
    \right] \ .
\end{equation}
Here, the disorder average is defined as ${\mathbb{E}[\cdots]
    =
    \int d\boldsymbol{J}\,
    \mathcal{P}(\boldsymbol{J})\,
    [\cdots]}$,
where $\boldsymbol{J}$ denotes the full set of binary couplings $J_{ij}$ defining the disordered Hamiltonian in Eq.~\eqref{eq:ham_model}. The probability distribution takes the form $\mathcal{P}(\boldsymbol{J}) = \prod_{ij} P(J_{ij})$,
with $P(J_{ij})$ defined in \cref{eq:prob_couplings}. In practice, the expectation value $\mathbb{E}[\cdots]$ in Eq.~\eqref{eq:loss} is approximated by an average over $\mathcal{R}$ disorder realizations sampled from $\mathcal{P}(\boldsymbol{J})$, with $\mathcal{R}$ typically of the order of hundreds. The variational parameters $\theta$ are optimized within the FNQS framework~\cite{rende2025foundation}. For each disorder realization, the variational energy is evaluated using Variational Monte Carlo~\cite{becca2017}. In the following, to simplify the notation, we denote by $\braket{\cdots}_{\boldsymbol{J}}$ the expectation value over the variational state $|\psi_{\theta}(\boldsymbol{J})\rangle$ at fixed disorder realization $\boldsymbol{J}$.
 
The key feature of the FNQS approach is that, at no additional computational cost compared to the single-instance case, a variational state parametrized through a Vision Transformer (ViT) architecture~\cite{viteritti2023prl, viteritti2024shastry, rende2024finetuning, sprague2024variational, rende2024stochastic} can be optimized simultaneously across multiple disorder realizations (refer to \textit{Supplementary Information} for additional details about the variational Ansatz)~\cite{rende2025foundation}.
This approach enables the efficient evaluation of disorder-averaged observables within a single simulation, substantially improving the scalability of variational methods for disordered systems. Remarkably, the accuracy of the model exhibits minimal degradation with increasing numbers of disorder realizations. Even when trained across hundreds of distinct disorder instances, the FNQS yields observable estimates that closely match those obtained from training on individual realizations with the same architecture (see~\cref{sec:ed} of the \textit{Methods}). \\

\subsection{Order parameters} 
To investigate the fate of conventional magnetic order at intermediate disorder strengths $0 < p < 1$, we compute the magnetic order parameters associated with the two limiting cases of the model: ferromagnetic order at $p = 0$ and antiferromagnetic (Néel) order at $p = 1$. The magnetic order parameters are defined as averages over the disorder distribution: ${\mathcal{M}^2_{\text{Ferro}} =  \mathbb{E} \left[m^2_{\text{Ferro}}(\b{J})\right]}$ and $\mathcal{M}^2_{\text{Néel}} = \mathbb{E} \left[ m^2_{\text{Néel}}(\b{J})\right]$, where, for a fixed realization $\b{J}$, the ferromagnetic order parameter is given by $m^2_{\text{Ferro}}(\b{J}) = C(0,0; \b{J}) / L^2$ while the Néel order parameter is $m^2_{\text{Néel}}(\b{J}) = C(\pi,\pi; \b{J}) / L^2$. Both quantities are computed from the spin structure factor, defined as $C(\boldsymbol{k}; \b{J}) = \sum_{\b{r}} e^{i \boldsymbol{k} \cdot \b{r}} \braket{ \hat{\boldsymbol{S}}_{\b{0}} \cdot \hat{\boldsymbol{S}}_{\b{r}} }_{\b{J}}$, where $\braket{ \hat{\boldsymbol{S}}_{\b{0}} \cdot \hat{\boldsymbol{S}}_{\b{r}} }_{\b{J}}$ is average over all the translations of the lattice (refer to \cref{sec:ed} of the \textit{Methods} for additional information). At the two limiting values of the disorder parameter $p = 0$ and $p = 1$, the distribution $\mathcal{P}(\b{J})$ becomes a delta function. Specifically, in the pure ferromagnetic case ($p = 0$), the ground state is fully polarized, and the order parameter can be computed exactly as ${\mathcal{M}^2_{\text{Ferro}} = \tfrac{1}{4} + \tfrac{1}{2L^2}}$~\footnote{By definition, $C(0,0) = \braket{ \hat{\boldsymbol{S}}^2 }$, where $\hat{\boldsymbol{S}}$ is the total spin operator. For a fully polarized state ${\braket{ \hat{\boldsymbol{S}}^2 } = S(S+1) = \tfrac{L^2}{2}(\tfrac{L^2}{2} + 1)}$, from which the order parameter follows.}. In the antiferromagnetic limit ($p = 1$), the model reduces to the standard square-lattice Heisenberg antiferromagnet, for which the Néel order parameter remains finite in the thermodynamic limit, with a known value of $\mathcal{M}_{\text{Néel}} = 0.3070(3)$~\cite{sandvik1997, calandra1998}.

In the interesting intermediate regime $0 < p <1$, no results are available from other numerical methods aside from exact diagonalization, which we perform up to $6 \times 6$ clusters (see \cref{sec:ed} of the \textit{Methods}). In the left panel of Fig.~\ref{fig:order_params}, we show the behavior of the magnetic order parameters as a function of $p$, for system sizes ranging from $L = 4$ to $L = 14$. Each data point corresponds to a single simulation that averages over $\mathcal{R} = 600$ disorder realizations. Then, for each value of $p$, we perform a finite-size extrapolation in $1/L$ to estimate the magnetic order parameters in the thermodynamic limit (see \cref{sec:extrapolations} of the \textit{Methods} for additional details). Our extensive numerical calculations identify a region in the phase diagram $0.2 \lesssim p \lesssim 0.8$ where both magnetic order parameters vanish in the thermodynamic limit.

When both magnetic order parameters vanish, it is still possible for the system to exhibit spin-glass order. In a spin glass phase, each spin is aligned along some preferred random direction, which is strongly realization-dependent. To capture the SG order, we examine the long-distance behavior of the correlation function squared, see, for example, Ref.~\cite{baity2015inherent}. We refer to this quantity as the Edwards-Anderson (EA) order parameter, denoted here as $Q$, and defined as:
\begin{equation}\label{eq:overlap_order_par}
    Q^2 = \mathbb{E} \left[\frac{1}{L^4}\sum_{\b{r},\b{r'}} \braket{\hat{\boldsymbol{S}}_{\b{r}} \cdot \hat{\boldsymbol{S}}_{\b{r'}}}_{\b J}^2 \right] \ ,
\end{equation}
where $\b{r}$ and $\b{r'}$ run over all lattice sites. The value of the EA order parameter is $Q_0^2 = S^4/3$ for a product state of classical, randomly oriented spins of the form $\prod_i \ket{\boldsymbol{S}_i}$, with $\boldsymbol{S}_i$ uniformly distributed on the Bloch sphere~\footnote{In fact, 
\begin{equation}
    {\frac{1}{L^4}\sum_{i,j}\sum_{\alpha,\beta}S^\alpha_i S^\alpha_jS^\beta_i S^\beta_j=S^4\sum_{\alpha,\beta} \frac{1}{3}\delta_{\alpha\beta}\frac{1}{3}\delta_{\alpha\beta}=\frac{S^4}{3}}
\end{equation}}. Any reduction from the classical value, i.e.~$Q^2<Q_0^2$, means that the fluctuations (quantum or thermal) reduce the persistence of the spins in the chosen random directions. 

\begin{figure}
    {\includegraphics[width=\columnwidth]{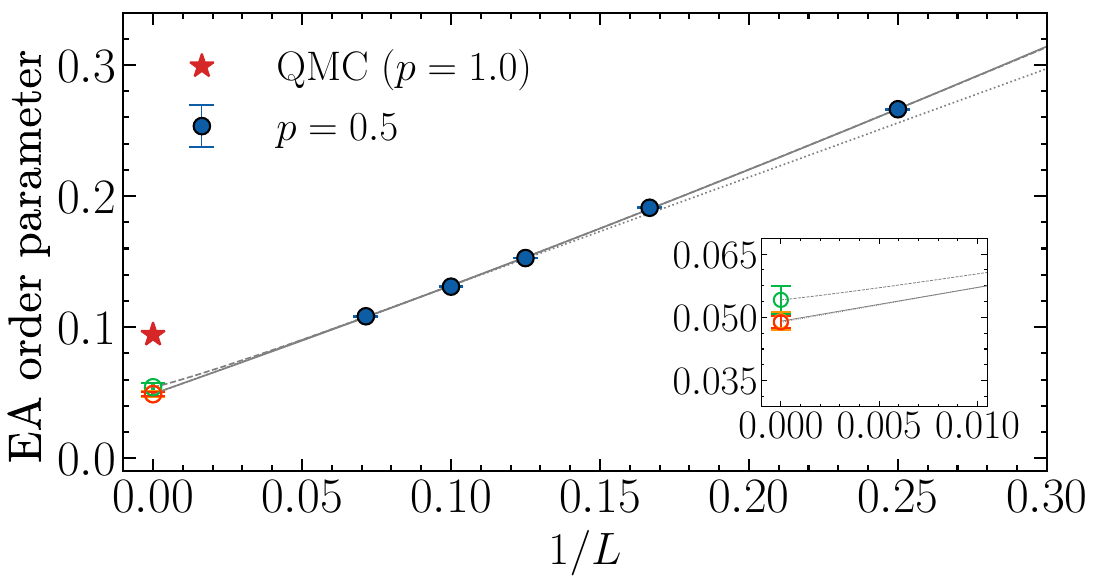}}
        \caption{Finite-size scaling of the Edwards-Anderson order parameter $Q$ [see Eq.~\eqref{eq:overlap_order_par}] as a function of the inverse system size $1/L$ for $L = 4$ to $L = 14$ at $p = 0.5$. Quadratic (solid line and yellow circle) and power-law (dashed line and green circle) fits are performed over all data points, while a linear fit (dotted line and red circle) is applied to the last three points. \textbf{Inset:} Zoom of the extrapolations close to $1/L \to 0$. The error bars of the  extrapolated values in the thermodynamic limit are estimated via a resampling technique with Gaussian noise. The red star indicating the extrapolated value at ${p = 1.0}$ is showed for comparison.}
        \label{fig:q_05}
\end{figure}

In Fig.~\ref{fig:q_05} we show the finite-size behavior of the EA parameter $Q$ as a function of the inverse system size $1/L$ for the challenging case $p = 0.5$.
To extrapolate to the thermodynamic limit, we consider different fitting forms, which give consistent estimates of the extrapolated value. In particular, the largest system sizes are well described by a linear scaling in $1/L$.
This behaviour is motivated by analogy with the scaling behavior of the other order parameters and is further supported by spin-wave theory. Indeed, in the clean antiferromagnetic limit ($p=1$), the finite-size correction to the Edwards-Anderson order parameter can be computed analytically in linear spin-wave theory and is found to scale as $1/L$. Moreover, the real-space spin-wave calculation performed in the disordered case displays the same leading behavior (see \emph{Supplementary Information}), providing support for the extrapolation procedure adopted here. 
By performing different finite-size extrapolations, we find strong numerical evidence that $Q$ remains finite at $p = 0.5$, with an extrapolated value estimated as $Q = 0.049(3)$. We repeat the analysis for different values of $p$, reporting the results in the right panel of Fig.~\ref{fig:order_params}. Extrapolating to the thermodynamic limit, we find that $Q$ remains finite throughout the intermediate region $0.2 \lesssim p \lesssim 0.8$, where both magnetic order parameters vanish (see \cref{sec:extrapolations} of the \textit{Methods} for additional details about the extrapolations). These results rule out disordered phases such as quantum spin liquids or valence-bond solids, both characterized by a vanishing EA order parameter, and provide compelling numerical evidence for the existence of a quantum spin-glass phase extending over a broad portion of the phase diagram. Although we cannot entirely exclude the presence of narrow intermediate regions near the magnetic transitions where $Q$ might vanish (in particular close to the antiferromagnetic phase transition), our findings strongly support that at least a quantum spin-glass phase is realized. Moreover, in the \textit{Supplementary Information}, we provide evidence that our approach can also describe cases in which both the magnetic order parameter and the EA order parameter vanish in the thermodynamic limit, as expected for states compatible with a quantum spin liquid~\cite{BeccaGutz2013, nomuraimada2021}. This benchmark shows that the finite value of the EA order parameter found in the disordered Heisenberg model is not an artifact of the variational method.

\subsection{Semiclassical analysis} 
The variational results obtained with the FNQS framework can be compared against a large-spin analysis by computing the leading order corrections in the inverse spin representation $1/S$ to the correlation functions and EA order parameter and then evaluating them in the fully quantum regime $S=\tfrac12$. This approach also provides physical insight into the QSG phase. The semiclassical approximation is based on the standard Holstein-Primakoff (HP) bosonic representation of the spin-$S$ operators \cite{Auerbach1994,blaizot_ripka_1986}. Spin waves in disordered systems have been extensively studied in the past \cite{Sherrington_1977, Takayama_1978, Ching_1981, Barnes_1981, Bray1981}; more recently, they have found application in the study of disordered superconductors~\cite{Cea2014,Menu2020} and of the one-dimensional version of the model under study~\cite{fava2024}. Building upon these results, we carry out the full computation of the correction to the spin glass order parameter in the spin wave approximation. 

\begin{figure}
    {\includegraphics[width=\columnwidth]{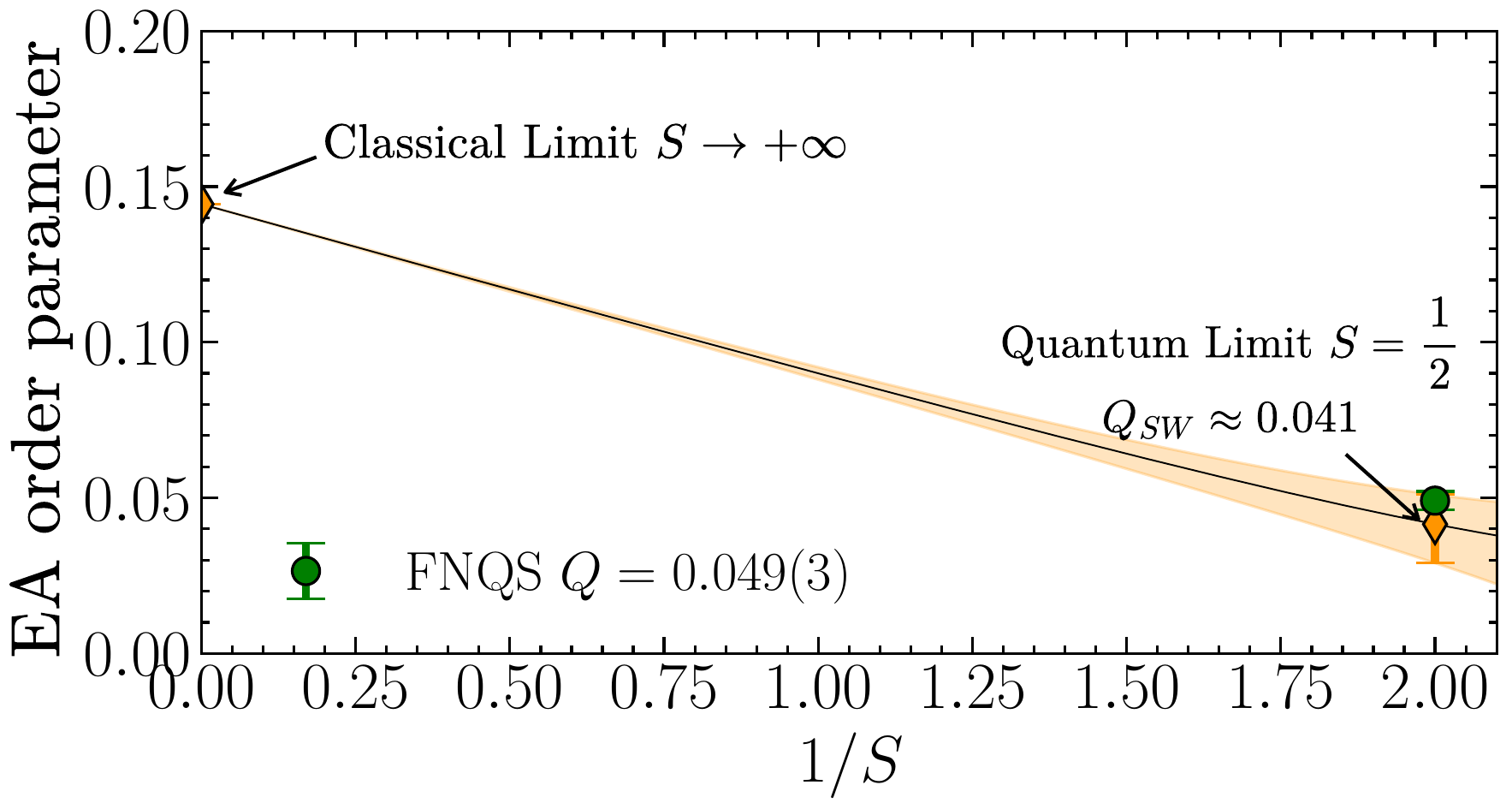}}
        \caption{ 
        Semiclassical prediction for the Edwards-Anderson order parameter in the thermodynamic limit as a function of the inverse spin representation $1/S$ at $p=0.5$. On the vertical axes we plot $Q_{\text{SW}}$ [see Eq.~\eqref{eq:Qlsw}] divided by $4S^2$ to have a comparison with the FNQS result at $S=\tfrac{1}{2}$ and at the same time with the classical value $Q_0 / (4S^2) = 0.1443$ (see main text for the definition). The semiclassical correction is obtained through a non-interacting spin wave-theory computation (see \emph{Semiclassical analysis} and the \emph{Supplementary Information} for details). For $S=\tfrac{1}{2}$ the value of the Edwards-Anderson order parameter is compatible with the value obtained for the fully quantum model with the FNQS, within the error bars.}
        \label{fig:semiclassical}
\end{figure}

For a classical disordered state, first a local change of reference is applied in order to align all spin operators on a fixed direction; then, the rotated spin operators are written in terms of bosonic creation and annihilation operators $\hat{a}_i,\hat{a}_i^\dagger$ satisfying canonical commutation relations $[\hat a_i,\hat a_j^\dagger]=\delta_{ij}$. The number operator $\hat a_i^\dagger \hat{a}_i$ encodes the reduction of the local magnetization compared to the classical Heisenberg vector $\boldsymbol{S}_i$. We truncate the HP representation obtained starting from the classical minimum at leading order in $1/S$. This yields a spin-wave correction to the classical energy $E_0$, such that the Hamiltonian takes the form $\hat{H}/S^2 =E_0 +\hat{H}_{\text{SW}}/S + o(1/S)$, with
\begin{equation}
    \hat{H}_{\text{SW}} = \sum_{\langle ij \rangle} \hat{a}_i^\dagger A_{ij} \hat{a}_j + \frac{1}{2} \sum_{\langle ij \rangle} \left( \hat{a}_i^\dagger B_{ij} \hat{a}_j^\dagger + \hat{a}_i B^*_{ij} \hat{a}_j \right) \; .
\end{equation}
The definition of the matrices $A,B$ is given in the \textit{Supplementary Information}.
This Hamiltonian breaks translational invariance and does not conserve particle number, so it must be diagonalized via a Bogoljubov transformation.
The ground state spin-spin correlation function yields a $1/S$ correction to the classical spin-spin correlation function:
\begin{equation}
\label{eq:corrSW}
\bra{0}\hat{\boldsymbol{S}}_i\cdot \hat{\boldsymbol{S}}_j\ket{0}= \boldsymbol{S}_{cl,i} \cdot \boldsymbol{S}_{cl,j} + \frac{1}{S} c^{(1)}_{ij} + o\left(\frac{1}{S}\right) \; ,
\end{equation}
where the expression of $c^{(1)}_{ij}$ is given in the \textit{Supplementary Information}. In the previous expression, $\ket{0}$ represents the Bogoljubov vacuum state, that is the state in which each spin wave is populated by zero quasiparticles. In particular, in order to decouple global rotations from the relative oscillations of the spins, we choose $\ket{0}$ such that the macroscopical operators implementing global spin rotations can be considered as \emph{c}-numbers and set to zero. 

By substituting Eq.~\eqref{eq:corrSW} into the definition of the EA parameter  [see Eq.~\eqref{eq:overlap_order_par}], we obtain 
\begin{equation}
\label{eq:Qlsw}
    Q_{\text{SW}}^2 = Q_0^2 + \Delta Q^2 \; ,
\end{equation}
where $\Delta Q^2$ is the leading order correction, whose expression can be found in the \textit{Supplementary Information}. In Fig.~\ref{fig:semiclassical} the semiclassical prediction to the EA order parameter is plotted as a function of $1/S$. The value at $S=\tfrac12$ is in agreement with the prediction of the FNQS within the error bars. A similar computation has been performed for other values of $p$ and the results are shown in the right panel of Fig.~\ref{fig:order_params} (see orange rhombi). Again, the semiclassical computation corroborates the FNQS prediction for the fully quantum model.

\section{Discussion} 
We explored the phase diagram of the two-dimensional Heisenberg model with binary disorder in the nearest-neighbor couplings, presenting compelling evidence for a quantum spin glass phase in the thermodynamic limit. Methodologically, we employed a novel variational framework based on FNQS~\cite{rende2025foundation}, which proves particularly effective for disordered quantum systems. Unlike conventional approaches, FNQS mitigate the high computational cost associated with averaging over many disorder realizations. We supported the numerical result by performing a semiclassical expansion around the classical minimum energy configuration. 

This work marks an important advance in the study of disordered two-dimensional quantum magnets, both conceptually and methodologically. On the conceptual side, it provides the first evidence, based on large-scale simulations, that the disordered Heisenberg model can host a quantum spin-glass phase. On the methodological side, it shows the application of an efficient and broadly applicable computational framework for investigating this class of frustrated, disordered spin systems.

The semiclassical analysis presented here has been restricted to the computation of the spin-glass order parameter. However, the same framework can be extended to investigate the low-energy excitations of the quantum spin-glass phase. In a companion work~\cite{braccitestasecca2026}, some of us showed that the excitation spectrum of the two-dimensional quantum Heisenberg spin glass exhibits a rich phenomenology. While the low-energy sector is expected to be described by the universal Halperin-Saslow hydrodynamic modes \cite{Halperin_Saslow1977}, linear spin-wave theory alone fails to reproduce this behavior because disorder suppresses diffusion in two dimensions. 

A key ingredient of the semiclassical analysis is the computation of the classical states around which the spin-wave expansion is performed. The search for the energy minima reveals that the underlying classical model possesses a highly nontrivial energy landscape, characterized by an exponential proliferation of local minima \cite{braccitestasecca2026}. Such behavior is consistent with the glassy nature of the problem and indicates that, unlike the special case of planar Ising spin glasses \cite{Bieche_1980}, the optimization of continuous-spin glass models remains a difficult and largely unresolved problem \cite{Weigel2007,Weigel2008,baity2013critical,baity2015inherent,agrawal2026}. The efficiency of the FNQS framework therefore does not originate from a simplification of the underlying classical landscape, but rather from its ability to exploit common structures across many disorder realizations and directly target disorder-averaged properties.

This work could be naturally extended along some of the following directions. First, a more detailed characterization of the antiferromagnetic transition near $p \approx 0.8$ seems compelling. In related frustrated models, this transition has been argued to support the emergence of quantum spin-liquid states~\cite{nomuraimada2021, BeccaGutz2013, viteritti2024shastry}; a spin-liquid phase could account for a small intermediate region between the antiferromagnetic and spin-glass phase where the EA order parameter can vanish. Second, key properties of the QSG state should be characterized through both dynamical and static response probes, for instance by studying the real-time response to weak local perturbations and the resulting propagation of dynamical correlations, as well as the effect of an infinitesimal bias field on the QSG ground state. Another interesting direction would be to compute stiffness distributions obtained from twisted boundary conditions. Within the FNQS framework, the response to both bias fields and twisted boundary conditions could be accessed efficiently by treating the bias field or the twisting angles as additional inputs of the neural network, allowing one to optimize simultaneously over disorder realizations and external parameters. These questions are beyond the scope of the present work and are left for future studies.

\section{Methods}

\subsection{Comparison with Exact Diagonalization}\label{sec:ed}
\begin{figure}[t]
    \begin{center}
\centerline{\includegraphics[width=\columnwidth]{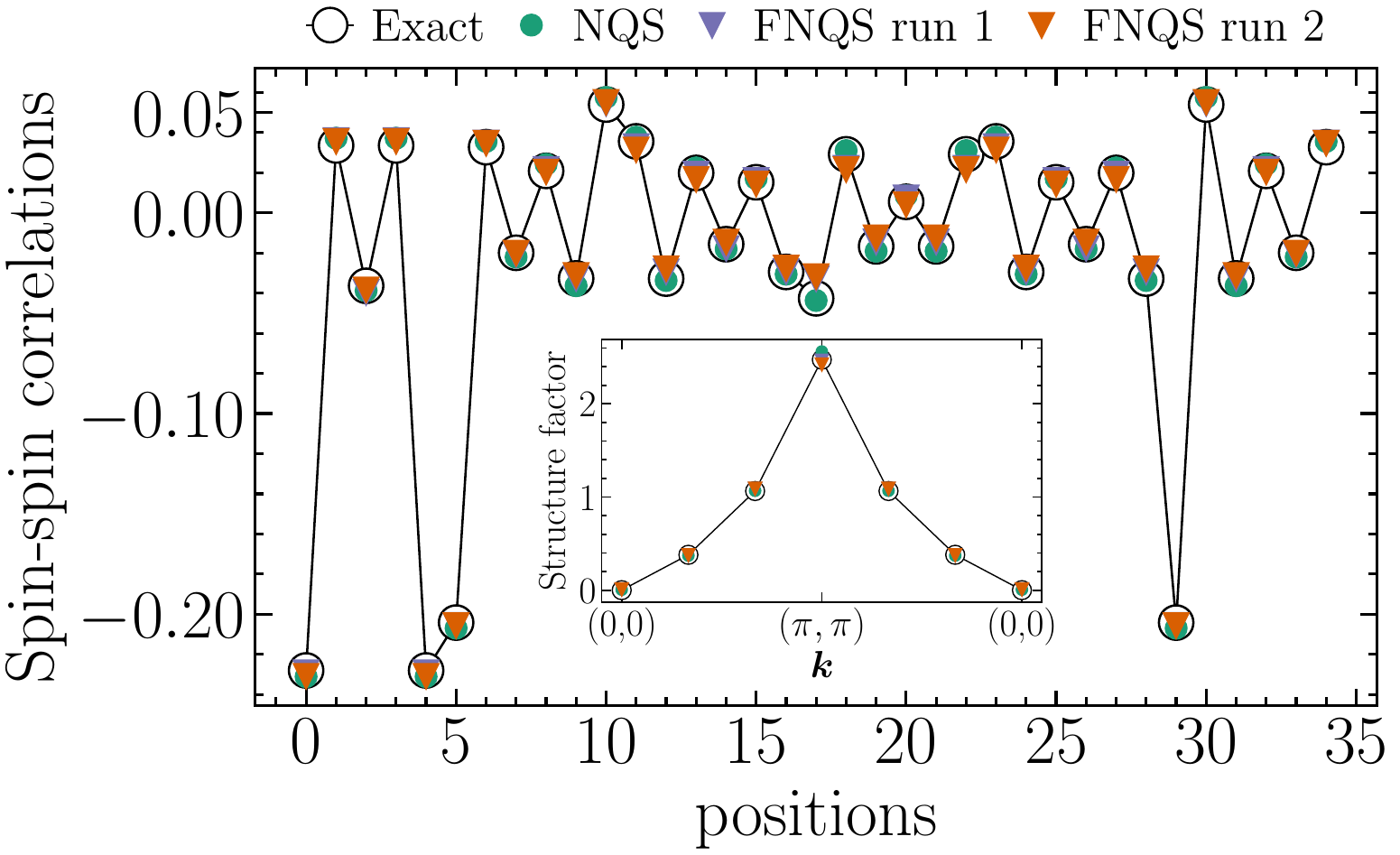}}
        \caption{\label{fig:spin_spin} 
Spin-spin correlations for a disorder realization with $p = 0.7$ on a $6 \times 6$ cluster. 
Spin-spin correlations for a representative disorder realization with $p=0.7$ on a $6\times6$ lattice. The two-dimensional cluster is unrolled using the row-major convention for visualization. Exact diagonalization results (empty black circles) are compared with variational calculations: an NQS optimized directly on the same disorder realization (green circles), and two independent FNQS runs trained on two different sets of $\mathcal{R}=600$ disorder realizations (purple and orange triangles), neither of which contains the realization shown here. The inset shows the corresponding spin structure factor along the path connecting $\boldsymbol{k}=(0,0)$ and $\boldsymbol{k}=(\pi,\pi)$.}
    \end{center}
\end{figure}

In \cref{fig:spin_spin}, we show the spin-spin correlations for a representative disorder realization with $p=0.7$ on a $6\times6$ cluster, which is the largest system size for which exact results from the Lanczos algorithm are available~\cite{sandvik2010computational}.

The simulations are performed on clusters with periodic boundary conditions in both directions. Therefore, the correlations are computed as
${\braket{\hat{\boldsymbol{S}}_{\boldsymbol{0}}\cdot\hat{\boldsymbol{S}}_{\boldsymbol{r}}}
=
\frac{1}{L^2}
\sum_{\boldsymbol{R}}
\braket{
\hat{\boldsymbol{S}}_{\boldsymbol{R}}
\cdot
\hat{\boldsymbol{S}}_{\boldsymbol{r}+\boldsymbol{R}}
}}$ ,
where $\boldsymbol{R}$ runs over all sites of the lattice and the sum is understood with periodic boundary conditions. In \cref{fig:spin_spin} we compare the exact data, shown as empty black circles, with different variational approaches.

First, we train a standard Neural-Network Quantum State (NQS)~\cite{carleo2017,lange2024}, based on a ViT architecture~\cite{viteritti2024shastry}, directly on this disorder realization. This corresponds to the usual setup in which a separate variational state is optimized for a fixed Hamiltonian. The resulting spin-spin correlations, shown as green circles in \cref{fig:spin_spin}, are in very good agreement with the exact data.

We then test the generalization properties of the FNQS. Specifically, we train two independent foundation models on two different training sets, each containing $\mathcal{R}=600$ disorder realizations. Importantly, the disorder realization used for the comparison with exact diagonalization is not included in either training set. We then evaluate both trained FNQS on this unseen realization and compare the corresponding spin-spin correlation functions with the exact results. Two aspects are relevant. First, the two variational states trained on different sets of disorder realizations give mutually consistent predictions. Second, their predictions agree very well both with the NQS trained directly on the target realization and with the exact Lanczos data, as shown by the purple and orange triangular markers in \cref{fig:spin_spin}.

Crucially, the spin structure factor, shown in the inset of \cref{fig:spin_spin}, is nearly identical for all variational methods. This shows that the magnetic order parameters, which are related to the structure factor at $\boldsymbol{k}=(0,0)$ and $\boldsymbol{k}=(\pi,\pi)$, are accurately reproduced; see \emph{Numerical Results}. The overall agreement with exact diagonalization confirms that FNQS can accurately describe individual disorder realizations and, importantly, can generalize to realizations that were not included during training~\cite{rende2025foundation}.

We emphasize that the standard optimization of a variational state for a single fixed disorder realization and the FNQS optimization over many disorder realizations have comparable computational costs. However, the FNQS provides a more general variational representation: once trained, it can be evaluated on unseen disorder realizations without retraining from scratch, making it particularly suitable for the study of disorder systems.

\begin{figure*}[t]
    \begin{center}
        \centerline{\includegraphics[width=2\columnwidth]{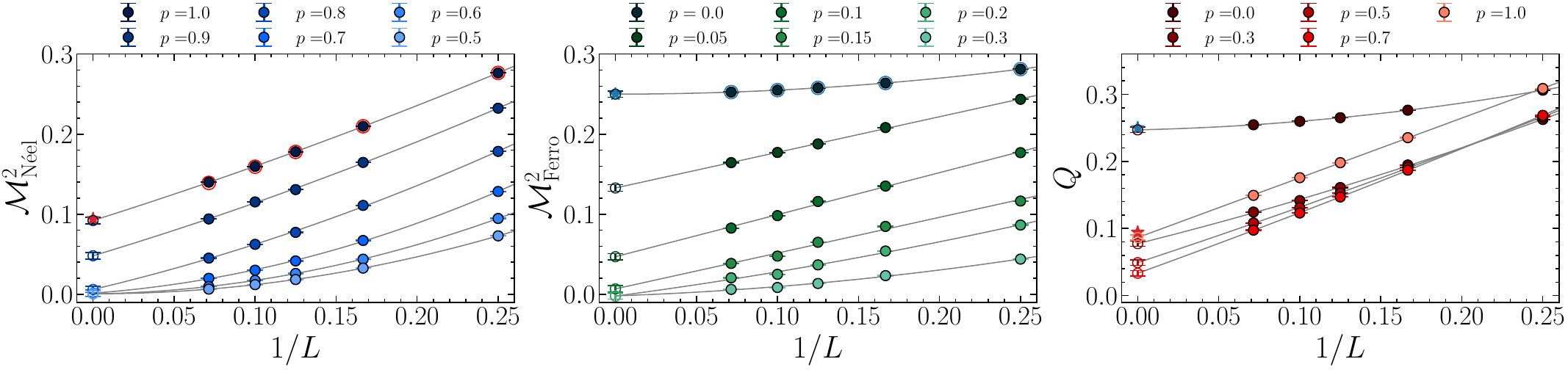}}
        \caption{Finite-size extrapolations of the order parameters as a function of $1/L$: the antiferromagnetic $\mathcal{M}^2_{\text{Néel}}$ (left panel), ferromagnetic $\mathcal{M}^2_{\text{Ferro}}$ (central panel), and Edwards-Anderson parameter $Q$ (right panel) order parameters, averaged over $\mathcal{R} = 600$ disorder realizations for system sizes ranging from $L = 4$ to $14$. The data for $L=4$ are obtained with exact diagonalization techniques. In the left panel, numerically exact Quantum Monte Carlo~\cite{sandvik1997} results for $p = 1.0$ are shown for finite sizes (red empty circles) and thermodynamic limit (red star). In the central panel, analytic results for $p = 0.0$ are shown for finite sizes (blue empty circles) and thermodynamic limit (blue star). In the right panel, thermodynamic-limit values of the Néel (red star) and ferromagnetic (blue star) order parameters are shown for comparison with the EA parameter $Q$ at $p = 1.0$ and $p = 0.0$, respectively. The error bars of the extrapolated values in the thermodynamic limit are estimated via a resampling technique with Gaussian noise. \label{fig:extrapolations}}
    \end{center}
\end{figure*}

\begin{figure*}[t]
    \begin{center}
        \centerline{\includegraphics[width=2\columnwidth]{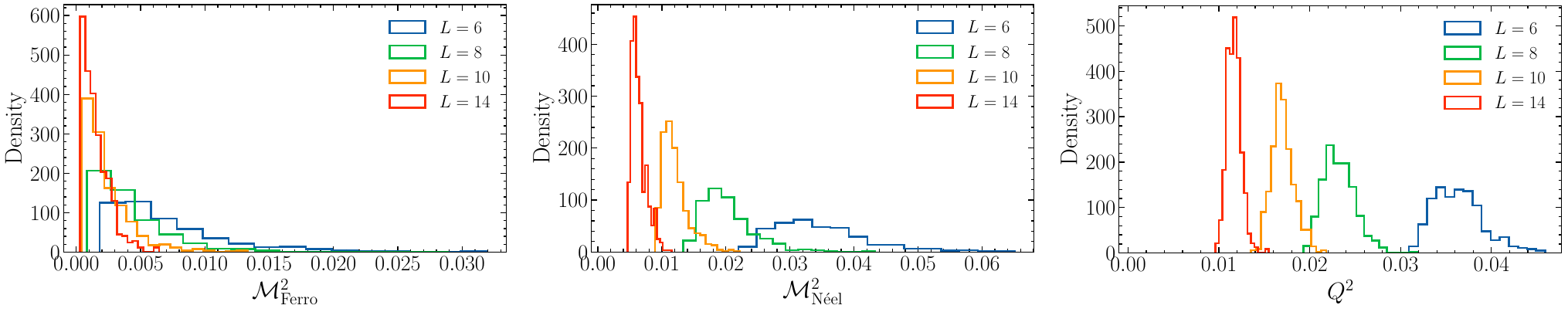}}
        \caption{Distributions of the order parameters $\mathcal{M}^2_{\text{Ferro}}$ (left panel), $\mathcal{M}^2_{\text{Néel}}$ (central panel), and $Q^2$ (right panel) for $p = 0.5$, considering system sizes ranging from $L = 6$ to $L = 14$. Each distribution is computed using $\mathcal{R} = 600$ independent disorder realizations.\label{fig:distributions_orderpar}}
    \end{center}
\end{figure*}

\subsection{Finite-Size Extrapolations to the Thermodynamic Limit}\label{sec:extrapolations}
In this section, we present the finite-size extrapolations of the three relevant order parameters: the antiferromagnetic order parameter $\mathcal{M}^2_{\text{Néel}}$, the ferromagnetic order parameter $\mathcal{M}^2_{\text{Ferro}}$, and the EA order parameter $Q$ (see \emph{Numerical Results} for their definitions). The extrapolations are shown as a function of inverse system size $1/L$ for system sizes ranging from $L=4$ to $L=14$. Each value of the order parameters is obtained by averaging over $\mathcal{R}=600$ disorder realizations. In the left panel of Fig.~\ref{fig:extrapolations}, we display the behavior of $\mathcal{M}^2_{\text{Néel}}$ for $p \in [0.4, 1.0]$. Notably, for $p_{\text{AFM}} \approx 0.8$, the extrapolated value vanishes in the thermodynamic limit, indicating the disappearance of antiferromagnetic long-range order. Additionally, we benchmark the variational results against Quantum Monte Carlo data~\cite{sandvik1994} for the unfrustrated case $p=1.0$, finding excellent agreement both at finite sizes and in the thermodynamic limit. The central panel of Fig.~\ref{fig:extrapolations} shows the extrapolation of the ferromagnetic order parameter $\mathcal{M}^2_{\text{Ferro}}$ for ${p \in [0.0, 0.3]}$. Here, for $p_{\text{FM}} \approx 0.2$, the extrapolated value also vanishes, signaling the suppression of ferromagnetic order. For $p=0$, we compare the variational results with the exact analytical prediction available for this limit (see Section \emph{Numerical Results}), again observing very good agreement. Finally, the right panel of Fig.~\ref{fig:extrapolations} reports the behavior of the EA parameter $Q$. For the limiting cases $p=0$ and $p=1$, we compare the extrapolated values of $Q$ with the corresponding magnetic order parameters, confirming consistency between the different observables. We also show the extrapolation of $Q$ for intermediate values $p=0.3$, $0.5$, and $0.7$, where both magnetic orders vanish in the thermodynamic limit. In contrast to the magnetic order parameters, $Q$ remains finite, suggesting the presence of a QSG phase in this parameter regime (refer to the \textit{Supplementary Information} for additional analysis on the EA order parameter in a quantum spin liquid state).

\subsection{Self-Averaging and Distribution of Order Parameters
}\label{sec:distributions}
In disordered systems, a central requirement for the validity of statistical predictions is the \textit{self-averaging} property of extensive observables \cite{Mezard87}. An observable is said to be self-averaging if its fluctuations across different disorder realizations vanish in the thermodynamic limit. This implies that ensemble-averaged quantities become representative of individual realizations as the system size increases. In Fig. ~\ref{fig:distributions_orderpar}, we analyze the distributions of the relevant order parameters for $p=0.5$, namely, the antiferromagnetic order parameter $\mathcal{M}^2_{\text{Néel}}$ (left panel), the ferromagnetic order parameter $\mathcal{M}^2_{\text{ferro}}$ (central panel), and the square of EA order parameter $Q^2$ (right panel), computed over $\mathcal{R}=600$ independent disorder realizations. The width of these distributions systematically decreases with increasing system size 
$L$, providing strong evidence of self-averaging. This property can be viewed as a central-limit-theorem-type narrowing, rather than as
evidence for additional nontrivial universal information encoded in the distribution.

We emphasize that this property is not only fundamental from a theoretical standpoint, but it also has important computational implications in our variational approach. In the FNQS framework, a single variational wave function is optimized simultaneously across many disorder realizations. The efficiency and accuracy of this global optimization strategy improve with system size in a self-averaging model: as the fluctuations between different realizations diminish, the structure of the ground states becomes more similar across the various disorder realizations. Consequently, the variational state is able to effectively capture the common features of the ground states in the disordered ensemble.

\begin{acknowledgments}
We thank A. Laio, F. Becca, 
S. Sachdev, and G. Parisi for useful discussions. 
R.R. and L.L.V. acknowledge the CINECA award under the ISCRA initiative for the availability of high-performance computing resources and support. The work of A.S.\ and J.N.\ was funded by the European Union--NextGenerationEU under the project NRRP Project ``National Quantum Science and Technology Institute" — NQSTI, Award Number: PE00000023, Concession Decree No.~1564 of 11.10.2022 adopted by the Italian Ministry of Research, CUP J97G22000390007. This work was in part supported by the Deutsche Forschungsgemeinschaft under grants SFB 1143 (project-id 247310070) and the cluster of excellence ct.qmat (EXC 2147, project-id 390858490). This work was supported by the Swiss National Science Foundation under Grant No. 200021\_200336. This research was also supported by SEFRI through Grant No. MB22.00051 (NEQS - Neural
Quantum Simulation).
\end{acknowledgments}

\section*{Data availability}
The architectures trained in this work are publicly available at \href{https://huggingface.co/nqs-models}{Hugging Face NQS models}.

\bibliography{ref}

\end{document}


\title{\textit{Supplementary Information of} Quantum Spin Glass in the Two-Dimensional Disordered Heisenberg Model via Foundation Neural-Network Quantum States}

\maketitle 

\section{Foundation Neural-Network Architecture}
The computational advantage of Foundation Neural-Network Quantum States (FNQS) is achieved by making the variational many-body wave function amplitudes $\psi_{\theta}(\b\sigma|\b J) = \langle \b \sigma | \psi_\theta (\b J ) \rangle$ explicitly dependent on the Hamiltonian couplings $\b{J}$. Here, $\b{\sigma}$ denotes a spin configuration on a two-dimensional square lattice of linear size $L$, with local spin variables $\sigma_{i} = \pm 1$ at each site $i$, where $i = 1, \dots, L^2$. 
The variational state $\psi_{\theta}(\b\sigma|\b J)$ is parametrized with a Vision Transformer (ViT) architecture~\cite{viteritti2023prl, viteritti2024shastry, rende2024finetuning, sprague2024variational, rende2024stochastic},
which processes a sequence of $n$ input vectors $\b{x}_1, \dots, \b{x}_n$, each in $\mathbb{R}^d$, where $d$ is a tunable embedding dimension. In the FNQS framework, the input to the neural network is built to incorporate the Hamiltonian couplings $\b J$ as input features of the neural network at the same level as the physical spin configurations $\b \sigma$~\cite{rende2025foundation}. Specifically, the spin configuration $\b\sigma$ is divided into $n$ non-overlapping patches of size $b^2$~\cite{viteritti2024shastry}, each embedded into $\mathbb{R}^{d/2}$ through a learnable linear transformation, producing the sequence $\b{x}^\sigma_k$ with $k=1,\dots, n$. At the same time, each spin $\sigma_{i}$ is associated with its two local couplings: the horizontal coupling $J_{i,i+1}$ and the vertical coupling $J_{i,i+L}$, assuming periodic boundary conditions. Using the same patching scheme, the sets of horizontal and vertical couplings are partitioned into patches and independently embedded into two sequences of vectors, $\b{x}^h_k$ and $\b{x}^v_k$ with $k=1, \dots, n$, each lying in $\mathbb{R}^{d/4}$. Finally, the input sequence of the model is obtained by concatenating the three embedding vectors at each patch location: $\b{x}_k \equiv \text{Concat}(\b{x}^\sigma_k, \b{x}^h_k, \b{x}^v_k) \in \mathbb{R}^d$. This set of $n$ vectors is processed by the ViT wave function, yielding as output a set of another $n$ vectors $\b{y}_k \in \mathbb{R}^d$. The scalar amplitude $\text{Log}[\psi_{\theta}(\b\sigma|\b J)]$ is obtained by summing over the output tokens, $\b{z} = \sum_{k=1}^n \b{y}_k$, and applying a complex-valued two-layer feedforward network with a single output neuron, as detailed in Refs.~\cite{viteritti2024shastry, rende2025foundation}.
The hyperparameters of the ViT architecture employed in the work are: $n_l = 6$ layers, $h = 14$ attention heads, and embedding dimension $d = 112$ (see Ref.~\cite{viteritti2024shastry} for more details about their role). The total number of variational parameters is $P=793240$ for $L = 6$ and $P=988120$ for $L = 14$. The FNQS state is optimized via Stochastic Reconfiguration (SR)~\cite{sorella1998, sorella2005, rende2024stochastic, chen2024empowering} for $10^4$ steps, using a number of samples of $M=6000$ configurations evenly distributed among $\mathcal{R}=600$ disorder realizations used for training. The learning rate follows a cosine decay schedule with an initial value $\eta = 0.09$, and a diagonal shift parameter of $\lambda=10^{-6}$ is used to regularize the inversion of the SR matrix (see Ref.~\cite{rende2024stochastic}). To accelerate convergence in highly frustrated regimes, we employ a transfer learning protocol: the optimized parameters at a given $p$ (e.g., $p=0.9$) are used as initialization for nearby values (e.g., $p=0.8$). We have verified that the final estimates of the order parameters are consistent with those obtained from longer simulations initialized from scratch.

\begin{figure}
    \begin{center}
    \centerline{\includegraphics[width=0.65\columnwidth]{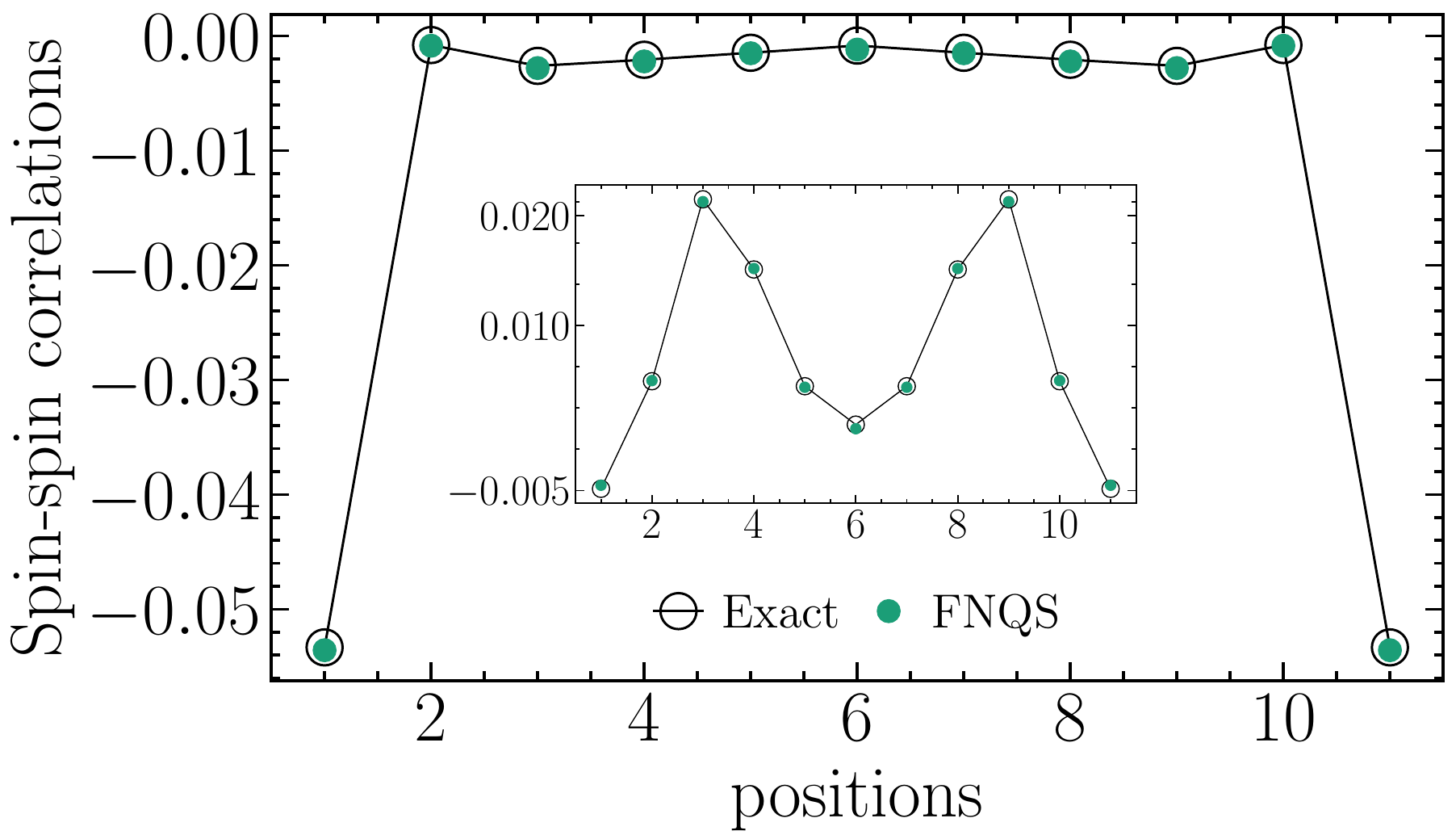}}
        \caption{Spin-spin correlations for the one-dimensional random-sign Heisenberg model [see Eq.~\ref{eq:heis_1d}] on a chain of length $L=12$. The FNQS is optimized simultaneously over $\mathcal{R}=512$ disorder realizations and compared with exact diagonalization. The main panel shows the disorder-averaged spin-spin correlations, while the inset shows the correlations for a single disorder realization not included in the FNQS training set.}
        \label{fig:spin_spin_heis_1d}
    \end{center}
\end{figure}

\section{One-dimensional Random-sign Heisenberg model}
In this Section, we benchmark our method on the one-dimensional random-sign Heisenberg model
\begin{equation}\label{eq:heis_1d}
    \hat{H} = \sum_{i=1}^N J_i \hat{\boldsymbol{S}}_i \cdot \hat{\boldsymbol{S}}_{i+1} \ ,
\end{equation}
where the couplings $J_i$ take the values $\pm1$ with equal probability. This model has recently attracted renewed interest and has been investigated using different numerical approaches, including DMRG~\cite{fava2024} and Quantum Monte Carlo~\cite{sibei2025}. As discussed in the main text, these studies reached different conclusions regarding the nature of the ground state. Here, we do not aim to settle this issue. Rather, we use small system sizes, for which exact diagonalization is available, to benchmark the FNQS approach and to test its generalization properties.

In Fig.~\ref{fig:spin_spin_heis_1d}, we compare spin-spin correlation functions (see definition in the main text) obtained with FNQS against exact diagonalization. The FNQS is optimized simultaneously over $\mathcal{R}=512$ disorder realizations. In the main panel, we compare the disorder-averaged correlations with the exact average obtained by diagonalizing the Hamiltonian over all disorder realizations. In the inset, we instead consider a single disorder realization that was not included in the training set and compare the FNQS prediction with the corresponding exact result. In both cases, we find very good agreement between the variational and exact results.

This benchmark shows that FNQS can accurately describe the one-dimensional random-sign Heisenberg model and can generalize to unseen disorder realizations. Based on these small-cluster benchmarks alone, we do not aim to make a quantitative comparison with Quantum Monte Carlo in one dimension, where the absence of a sign problem allows simulations of very large system sizes~\cite{sibei2025}. Nevertheless, neural-network states are known to be very accurate for sign-problem-free systems~\cite{carleo2017}, and the FNQS approach remains particularly promising because it provides an efficient way to treat disorder averages by optimizing simultaneously over many realizations. These results therefore indicate that FNQS may provide a useful complementary tool for future studies of this model, in particular for clarifying the nature of the ground state in regimes where different numerical approaches have led to apparently conflicting conclusions.

\begin{figure}
    \begin{center}
    \centerline{\includegraphics[width=0.7\columnwidth]{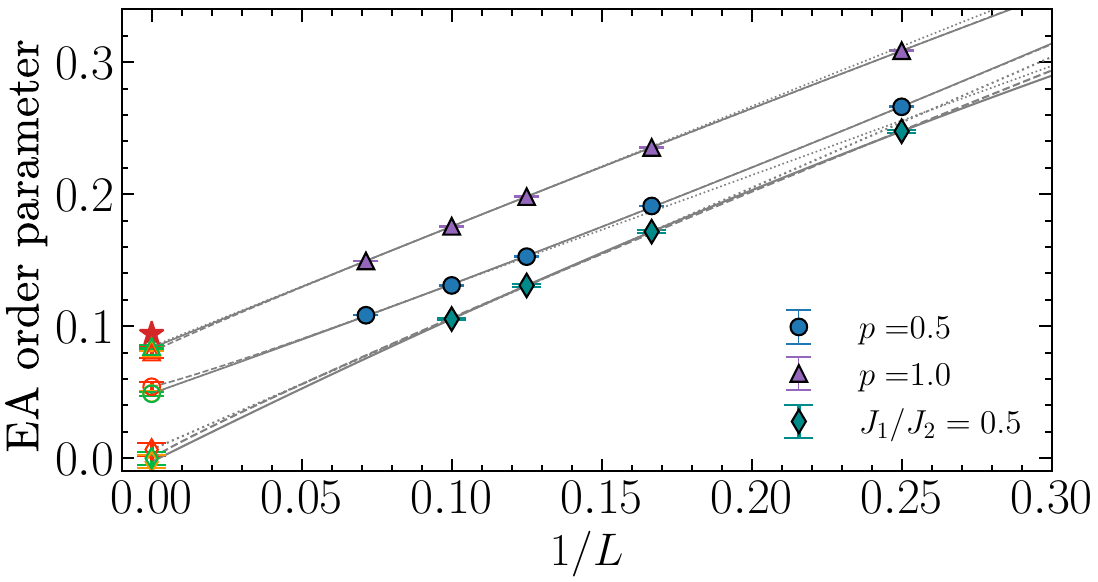}}
        \caption{Finite-size scaling of the Edwards-Anderson order parameter $Q$ as a function of the inverse system size $1/L$. 
Quadratic (solid line and yellow markers) and power-law (dashed line and green markers) fits are performed over all data points, 
while a linear fit (dotted line and red markers) is applied to the last three points. 
Results are shown for the disordered Heisenberg model at $p = 0.5$ (blue circles) and $p = 1.0$ (purple triangles), 
and for the $J_1$-$J_2$ Heisenberg model at $J_2/J_1 = 0.5$ (green diamonds), 
with system sizes ranging from $L = 4$ to $L = 14$ (and up to $L = 10$ for the $J_1$-$J_2$ Heisenberg model). 
Error bars of the extrapolated thermodynamic values are obtained via a resampling procedure with Gaussian noise. 
The red star marks the extrapolated value at $p = 1.0$, corresponding to the square of the N\'eel magnetization of the Heisenberg model ($J_2/J_1 = 0$) 
in the thermodynamic limit, shown here for comparison.}
        \label{fig:extrapolations_spin_liquid}
    \end{center}
\end{figure}

\section{Edwards-Anderson Order Parameter in the $J_1$-$J_2$ Heisenberg model}
In this Section, we use the $J_1$-$J_2$ Heisenberg model as a benchmark to test the behavior of the Edwards-Anderson order parameter $Q$ (see definition in the main text) in a system where numerical evidence indicates the absence of magnetic order in the thermodynamic limit. The model is defined on an $L \times L$ square lattice with periodic boundary conditions by the Hamiltonian
\begin{equation}\label{eq:J1J2_ham}
  \hat{H}
  =
  J_1\sum_{\langle i,j \rangle}
  \hat{\mathbf{S}}_{i}\cdot\hat{\mathbf{S}}_{j}
  +
  J_2\sum_{\langle\langle i,j \rangle\rangle}
  \hat{\mathbf{S}}_{i}\cdot\hat{\mathbf{S}}_{j} \, .
\end{equation}
We focus on the highly frustrated point $J_2/J_1=0.5$, where several numerical studies have reported the disappearance of N\'eel order and proposed a nonmagnetic ground state, possibly compatible with a gapless spin-liquid behavior~\cite{nomuraimada2021,BeccaGutz2013}. We stress, however, that the nature of the nonmagnetic phase of the $J_1$-$J_2$ Heisenberg model remains debated~\cite{capriotti2000, nomuraimada2021,BeccaGutz2013, chen2024empowering, DMRGSheng2014}, and that we do not use this model as a definitive or archetypal realization of an algebraic spin liquid.

Rather, the purpose of this comparison is methodological. We want to verify that the neural-network variational approach can also describe a case in which the magnetic order parameter vanishes and the Edwards-Anderson order parameter does not extrapolate to a finite value. This provides a useful complementary benchmark to the disordered Heisenberg model studied in the main text, where the magnetic order parameter vanishes while $Q$ remains finite in the thermodynamic limit.

For a translationally invariant periodic system, $Q$ can be written as $Q^2
=
\frac{1}{L^2}
\sum_{\boldsymbol{r}}
\langle
\hat{\boldsymbol{S}}_{\boldsymbol{0}}
\cdot
\hat{\boldsymbol{S}}_{\boldsymbol{r}}\rangle^2$. If the spin-spin correlations decay algebraically at long distances, $\langle
\hat{\boldsymbol{S}}_{\boldsymbol{0}}
\cdot
\hat{\boldsymbol{S}}_{\boldsymbol{r}}
\rangle
\sim
\frac{1}{|\boldsymbol{r}|^{\eta+z}}$, with $\eta+z>0$, then $Q$ vanishes in the thermodynamic limit. For the $J_1$-$J_2$ model at $J_2/J_1=0.5$, previous numerical studies reported algebraic correlations with $\eta+z \approx 1.5$~\cite{nomuraimada2021}, which are therefore consistent with $Q(L\to\infty)=0$.

In Fig.~\ref{fig:extrapolations_spin_liquid}, we compare the finite-size behavior of $Q$ in three representative cases. For the unfrustrated Heisenberg antiferromagnet, $J_2/J_1=0$, $Q$ extrapolates to a finite value related to the square of the N\'eel order parameter. For the frustrated $J_1$--$J_2$ model at $J_2/J_1=0.5$, $Q$ decreases with system size and is consistent with a vanishing thermodynamic-limit value. By contrast, for the disordered Heisenberg model at $p=0.5$, $Q$ remains finite while the magnetic order parameter vanishes. This comparison shows that the finite value of $Q$ found in the disordered case is not a generic artifact of the variational Ansatz, since the same numerical framework is able to reproduce a vanishing $Q$ in a nonmagnetic benchmark system.

\section{Semiclassical Expansion}
The semiclassical expansion has proven to be an efficient way to capture properties of quantum systems, being able to predict, for example, the difference between the quadratic/linear dispersion relation in one-dimensional ferromagnetic/antiferromagnetic spin-$1/2$ chains \cite{Dupuis_CM}. It relies on the Holstein-Primakoff (HP) mapping of a spin $S$ operator onto bosonic variables, which at the leading order in the classical limit $S \to \infty,~ \hbar \to  0$ simplifies to a quadratic (non-interacting) problem. In the following, we present the derivation of the procedure that allows for the computation of the quantum correction to the SG order parameter discussed in the main text (see \emph{Semiclassical Expansion}).

The procedure is divided into two steps: first, we look for the classical ground state, i.e.~the global minimum of the classical energy for a given coupling realization; then, we build up the HP machinery, which produces the quadratic bosonic Hamiltonian, describing quantum fluctuations around the classical configuration. The classical energy function is defined as
\begin{equation}
    E_{cl} = \sum_{ \langle i,j \rangle} J_{ij} \mathbf{S}_i \cdot \mathbf{S}_j \; ,
    \label{eq:classicalHam}
\end{equation}
where $\mathbf{S}_i$ is an $O(3)$ vector (i.e.~$|\mathbf{S}_i|^2=1$) and $J_{ij}=\pm 1$ a random sign nearest neighbors coupling term. The model is known to have a SG phase in $d=3$ on the cubic lattice \cite{Baity2015soft}, while in $d=2$, SG order is confined to $T=0$ by Mermin-Wagner-like arguments \cite{Kawamura03_2D, Fernandez1977}.

To find the classical ground state, we resort to a greedy single-spin update optimization algorithm, similar to the one presented e.g.~in Ref.~\cite{baity2015inherent}. For a given initial condition, the algorithm performs a deterministic energy minimization procedure, which crucially depends on the number of restarts: in general, for large system sizes $L$ and a reasonable number of restarts $N_r$, finding the global minimum of Eq.~\eqref{eq:classicalHam} is not guaranteed, though the occurrence of low energy minima is much larger then the occurrence of high energy minima \cite{braccitestasecca2026}. However, this is not too annoying for our purposes, since it can be checked \emph{a posteriori} that the outcome of the semiclassical computation does not depend on whether the global or another low energy local minimum has been selected. To this end, different classical minima have been considered and it has been checked that the semiclassical prediction on the SG order parameter are not correlated with the energy of the local minima. This is not surprising, since the semiclassical expansion is a perturbative computation that depends only on local quantities, and is consistent with the self-averaging behavior of the Hessian spectrum.

\section{The Holstein-Primakoff transformation}
A spin $S$ operator can be mapped onto a set of bosonic variables via the HP transformation defined as
\begin{equation}
\hat{S}^z_i = S - \hat{a}^\dagger_i \hat{a_i} \; , \quad
\hat{S_i}^+ = \sqrt{2S - \hat{a_i}^\dagger\hat{a_i}} \; \hat{a_i} \; , \quad
\hat{S_i}^- = \hat{a_i}^\dagger \sqrt{2S-\hat{a}^\dagger_i \hat{a_i}} \; ,
\end{equation}
where $\hat{a_i},\hat{a}^\dagger_i$ are bosonic annihilation/creation operators satisfying canonical commutation rules $[\hat{a}_i, \hat{a}_j^\dagger] = \delta_{ij}, \; [\hat{a}_i,\hat{a}_j] = [\hat{a}^\dagger_i,\hat{a}^\dagger_j] = 0$. This map preserves the spin algebra $[\hat{S}_i^\alpha,\hat{S}_j^\beta] = \delta_{ij}\epsilon_{\alpha \beta \gamma}\hat{S}^\gamma$, where $\alpha, \beta, \gamma = \{ x,y,z \}$ and $\hat{S}^\pm = \hat{S}^x \pm i \hat{S}^y$.

In the limit of large spin $S$ one obtains, at leading order, the HP map takes the following form:
\begin{equation}
\label{eq:standardHP}
    \hat{S}^z_i \sim S - \hat{a}^\dagger_i \hat{a}_i \; , \quad
    \hat{S}^y_i \sim -i\sqrt{\frac{S}{2}}(\hat{a}_i - \hat{a}^\dagger_i) \; ,\quad
    \hat{S}^x_i \sim \sqrt{\frac{S}{2}}(\hat{a}_i + \hat{a}^\dagger_i) \; .
\end{equation}
When applied to the study of quantum fluctuations around a classical minimum, the HP transformation has an intuitive interpretation: the operators $a,a^\dagger$ generate excitations in the plane orthogonal to the classical direction; the number of particle operator $\hat{n}_i \equiv \hat{a}^\dagger_i \hat{a}_i$ measures the deviation from the classical spin due to quantum fluctuation and the corresponding amplitude of fluctuation in the orthogonal plane. These excitations are quantized according to the possible values of the spin in the representation $S$ of the $SU(2)$ group and are known as \emph{spin waves} or \emph{magnons}.

\subsection*{Disordered Holstein-Primakoff}
For a spin-glass random configuration, or in general a configuration that is not invariant under any lattice symmetry (as instead a Neél state or a ferromagnetically ordered state), building the HP Hamiltonian is not straightforward. Consider a minimum energy configuration $\{\mathbf{S}\}$ of the $d=2$ disordered Heisenberg model, defined by Eq.~\eqref{eq:classicalHam}. The first step of our procedure consist in performing a local change of reference and rewriting the classical Hamiltonian as
\begin{equation} \label{eq:rotHam}
    E_{cl} = \sum_{ij} J_{ij} R_i^{\beta \alpha } R_j^{t, \alpha \gamma} \Sigma_i^\beta \Sigma_j^\gamma \; ,
\end{equation}
where $R^{\alpha \beta} S^\beta = \Sigma^\alpha$ and $\Sigma^\alpha = \hat{z}$, so that in the rotated configuration $\{\mathbf \Sigma \}$ all spins are all aligned along the $z$-axis, as for a pure ferromagnetic state. Let us also introduce the shorthand notation $\mathcal{R}_{ij}^{\alpha \beta} \equiv  R_i^{\alpha \gamma} R_j^{t,  \gamma \beta}$, where the superscript $t$ denotes matrix transposition and summation over repeated indices is assumed. 

Now, the HP transformation can be applied to the rotated spin operators at leading order in the large $S$ expansion, see Eqs.~\eqref{eq:standardHP}. In order for the quadratic term in the bosonic operators to be the coefficient of the order $O(1/S)$ in the expansion, let us divide the quantum Hamiltonian by $S^2$, so that it reads:
\begin{equation}
\label{eq:rotHam_quant}
    \hat H/S^2 = \sum_{ij} J_{ij} \mathcal{R}_{ij}^{\alpha \beta} \hat{\Sigma}_{ij}^{\alpha \beta} \; ,
\end{equation}
where the tensor $\hat{\Sigma}_{ij}^{\alpha \beta} = \hat{\Sigma}_i^\beta \hat{\Sigma}_j^\gamma / S^2$ has been defined. By computing all the components of the tensor $\hat{\Sigma}$, one obtains the following identities:
\begin{align} \label{eq:sigmatensor}
\hat \Sigma^{xx}_{ij} &= \frac{1}{2S} (\hat{a}_i + \hat{a}_i^\dagger)(\hat{a}_j + \hat{a}_j^\dagger) &
\hat \Sigma^{yy}_{ij} &= - \frac{1}{2S} (\hat{a}_i - \hat{a}_i^\dagger)(\hat{a}_j - \hat{a}_j^\dagger) \nonumber \\
\hat \Sigma^{zz}_{ij} &= 1 - \frac{1}{S}(\hat{a}_i^\dagger \hat{a}_i + \hat{a}_j^\dagger \hat{a}_j) &
\hat \Sigma^{xy}_{ij} &= - i\frac{1}{2S} (\hat{a}_i + \hat{a}_i^\dagger)(\hat{a}_j - \hat{a}_j^\dagger) \nonumber \\
\hat \Sigma^{yz}_{ij} &= -i\frac{1}{\sqrt{2S}} (\hat{a}_i - \hat{a}_i^\dagger) 
&
\hat \Sigma^{xz}_{ij} &= \frac{1}{\sqrt{2S}} (\hat{a}_i + \hat{a}_i^\dagger) \; .
\end{align}
Three terms can be recognized in the tensor $\hat{\Sigma}$: $\hat \Sigma = \hat \Sigma^{(0)} + \hat{\Sigma}^{(1)} + \hat{\Sigma}^{(2)}$. The first term contains no bosonic operators and returns the classical energy; the second term can be ignored because it is linear in the bosonic operators and therefore vanishes, as the expansion is around a minimum; the third term yields the quadratic spin-wave Hamiltonian. Then, the quantum Hamiltonian takes the form $\hat{H} / S^2 = E_{cl} + \hat{H}_{SW}/S + o(1/S)$, with $E_{cl} \sim O(1)$ and $\hat{H}_{SW}$ defined as
\begin{equation}
    \hat H_{SW} =  \sum_{ij} \hat{a}_i^\dagger A_{ij} \hat{a}_j + \frac{1}{2} \sum_{ij} \left( \hat{a}_i^\dagger B_{ij} \hat{a}_j^\dagger + \hat{a}_i B^*_{ij} \hat{a}_j \right) \; ,
    \label{eq:HSW}
\end{equation}
where the explicit expression of the matrix elements is, for $i\neq j$:
\begin{subequations}
    \begin{align}
        A_{ij} &= J_{ij} \mathcal{A}_{ij} = \frac{J_{ij}}{2}\left( \mathcal{R}_{ij}^{xx} + \mathcal{R}_{ij}^{yy}\right) + \frac{iJ_{ij}}{2} \left( \mathcal{R}_{ij}^{yx} - \mathcal{R}_{ij}^{xy}\right) \\
        B_{ij} &= J_{ij} \mathcal{B}_{ij} =  \frac{J_{ij}}{2} \left( \mathcal{R}_{ij}^{xx} - \mathcal{R}_{ij}^{yy}\right) + \frac{iJ_{ij}}{2} \left( \mathcal{R}_{ij}^{yx} +
        \mathcal{R}_{ij}^{xy}\right) \; ,
    \end{align}
\end{subequations}
and for the diagonal elements:
\begin{equation}
A_{ii} = -\sum_{j \in \partial i}J_{ij} \mathcal{R}_{ij}^{zz} \; ,\quad
B_{ii} = 0 \; .
\end{equation}

The spin-wave Hamiltonian in Eq.~\eqref{eq:HSW} can be rearranged in a more compact expression by introducing the $2 N$ dimensional vector $\hat \alpha = (\hat a_1, \cdots,\hat a_N, \hat a^\dagger_1, \cdots, \hat a_N^\dagger)^t$, with $N=L^2$, and its adjoint $\hat \alpha^\dagger \gamma = 
\hat \alpha^t $, where $\gamma = \begin{bmatrix}
0 & 1_N  \\
1_N & 0  \\
\end{bmatrix}$.
By introducing the $2N \times 2N$  coefficients matrix
$M = \begin{bmatrix}
A & B  \\
B^* & A^* \\
\end{bmatrix}$
the Hamiltonian can be expressed as  
\begin{equation}
   \hat H_{SW} = \frac{1}{2}\hat \alpha^\dagger M \hat \alpha  -\frac{1}{2} \mathrm{t
    r} A \; .
    \label{eq:defH}
\end{equation}

\section{Bogoljubov theory} 
The quadratic Hamiltonian in Eqs.~(\ref{eq:HSW},\ref{eq:defH}), for some matrices $A$ and $B$, defines a completely generic non-interacting problem, in which both translational invariance and particle number conservation are absent. To solve the problem, one needs to find a linear transformation $T \in M_{2N \times 2N}(\mathbb{C})$, not unitary in general, which diagonalizes the quadratic form $M$, while preserving the bosonic commutation rules on the quasiparticle operators $\beta= T \alpha$, where $\beta^t = ( \beta_1 , \cdots , \beta_N, \beta^\dagger_1, \cdots, \beta^\dagger_N) $. This procedure is known as generalized Bogoljubov transformation \cite{Bogo1947theory, fetterwalecka, blaizot_ripka_1986, COLPA1978}.

The bosonic commutation relations for the Bogoljubov quasiparticles $\beta$ can be written as
\begin{equation}
    [\beta, \beta^\dagger ] = \eta \; ,
\end{equation}
where $\eta$ is the $2 N$-dimensional metric form: $\eta=\text{diag}(1,1,\cdots,-1,-1, \cdots)$. 
This implies that the matrix $T$ satisfies 
\begin{equation}
    T\eta\gamma T^t = \eta \gamma \; ,
\end{equation}
meaning that $T$ is an element of the symplectic group $SP(2N,\mathbb{C})$. The requirement of unitarity $T = \gamma T^* \gamma$ combined with the previous condition gives 
\begin{equation}
    T \eta T^\dagger \eta = 1_{2N} \; .
\end{equation}
The last equation can be used to write an explicit form for the matrix $T$ and its inverse
\begin{equation}
T = \begin{bmatrix}
X^* & -Y^*  \\
-Y & X  \\
\end{bmatrix} \qquad
 T^{-1} = \begin{bmatrix}
X^t & Y^\dagger \\
Y^t
& X^\dagger  \\
\end{bmatrix} \; ,
\label{eq:BogandInv}
\end{equation}
where, as it will be clear in a while, the matrices $X$ and $Y$ contain the components of the eigenvectors of the matrix $\eta M$. From the previous expressions, a more explicit form of the canonicity conditions can be derived, which is the straightforward generalization of the translational invariant case, appearing in the theory of superconductors \cite{Bogo1947theory}:
\begin{equation}
    \begin{split}
    XX^\dagger - YY^\dagger = 1_N  \qquad X^\dagger X - Y^t Y^* = 1_N \\
    X Y^t -Y X^t = 0_N \qquad X^t Y^*  -Y^\dagger X=0_N  \; .      
    \end{split}
\end{equation}

By plugging the definition of the Bogoljubov transformation in Eq.~\eqref{eq:defH}, one finds
\begin{equation}
    \hat H_{SW} = \frac{1}{2} \beta^\dagger \eta T \eta M T^{-1} \beta - \frac{1}{2} \text{tr} A \; .
    \label{eq:Bogo-Ham generic}
\end{equation} 
Therefore, the solution of the problem reduces to finding the eigenvalues and eigenvectors of the non-Hermitian matrix $\eta M$, also called \textit{dynamical matrix} \cite{Xiao2009}. It is possible to show that if the matrix $M$ is positive definite, the spectrum of $\eta M $ is real \cite{blaizot_ripka_1986}. This is the physically relevant case, since the spin wave Hamiltonian is obtained by expanding around a classical minimum.

The matrix $\eta M$ is not hermitian, so it is not guaranteed that the algebraic dimension of a degenerate eigenspace --- the number of degenerate eigenvalues --- is equal to the geometric dimension of such eigenspace --- the number of linearly independent eigenvectors.  In a disordered system, typically there are no degenerate finite eigenvalues; however, the problem defined by Eq.~\eqref{eq:defH}
is characterized by the presence of zero modes associated to global spin rotations. We find that the degenerate kernel of the matrix $\eta M$ is a defective eigenspace: this prevents us to find the components of the eigenvectors corresponding to the global rotations within our numerical procedure. However, the overlap is a rotationally invariant quantity, so it does not depend on these modes. Let us denote with $\hat{\mathcal{B}}_n$ the zero modes of the theory, even if we are not able to compute their components numerically. In order to remove the contribution of these modes from the overlap correction, the ground state  $\ket{0}$ of the Bogoljubov theory can be chosen in such a way that $f( \hat{\mathcal{B}}_n ) \ket{0} = 0$, for any function $f$ such that $f(0) = 0$. In other terms, we choose the quasiparticle vacuum to be a coherent state of  the zero modes and consider the operators $\hat{\mathcal{B}}_n$ as \emph{c}-numbers, whose value can be set to 0 without loss of generality. This procedure is justified by the expectation that these operators are macroscopical, since they implement global rotations. In fact, the error committed is proportional to $1/N$, as we have carefully checked. 

Finally, on a more technical side, the numerical problem of dealing with vanishing eigenvalues within machine precision has been tackled by adding to the Hamiltonian a small symmetry breaking field on the first two sites, which slightly lifts the kernel eigenvalue. We performed a careful study of the dependence on the external field, showing that there is a range of values of the fields where the semiclassical prediction on the overlap correction does not depend on the value of the field and extrapolates to the correct result in the thermodynamic limit.

\subsection*{Two point correlation function and the order parameter quantum correction.}

In order to compute the effect of quantum fluctuations on the overlap order parameter in the semiclassical theory, one is interested in computing the leading order correction in the $1/S$ expansion to the 2-point correlation function, which enters the definition of $Q^2$:
\begin{equation}
    Q^2 = \frac{1}{L^4} \sum_{ij}^n C_{ij}^2 \; , 
\end{equation}
where $C_{ij} \equiv \langle \hat{\mathbf{S}}_i\cdot \hat{\mathbf{S}}_{j}\rangle$. It is convenient to use the following expression of the spin operator: $\hat{\mathbf{S}} = \mathbf{S}_{cl} + \delta \hat{\mathbf{S}}/S$, with $|\mathbf{S}_{cl}|^2=1$, and an equivalent one for the rotated spin operator $\hat{\mathbf{\Sigma}}$, for which $\mathbf{\Sigma}_{cl} = \hat{z}$. In the bosonic HP representation, the leading order quantum fluctuation is given by 
\begin{equation}
    \delta \hat{\mathbf{\Sigma}}/S = \begin{pmatrix}
        \frac{a^\dagger + a}{\sqrt{2 S}} \\
        -i \frac{a - a^\dagger}{\sqrt{2 S}} \\
        -\frac{a^\dagger a}{S}
    \end{pmatrix} = \frac{1}{\sqrt{S}} \delta  \hat{\mathbf{\Sigma}}_{1/2} + \frac{1}{S} \delta  \hat{\mathbf{\Sigma}}_{1} \; ,
\end{equation}
with $\delta  \hat{\mathbf{\Sigma}}_{1/2}$ and $\delta  \hat{\mathbf{\Sigma}}_{1}$ of order $O(1)$. In the previous expression, we have extracted the dependence on $S$ and separated the $x$ and $y$ components, which are of order $1/\sqrt{S}$, from the $z$ component, which is of order $1/S$. The quantum correlation function has the following expression:
\begin{equation}
    C_{ij} = \mathbf{S}_{cl,i}\cdot \mathbf{S}_{cl,j} + \mathbf{S}_{cl,i}\cdot \langle \delta \hat{\mathbf{S}}_j /S \rangle + \mathbf{S}_{cl,j}\cdot \langle \delta \hat{\mathbf{S}}_i / S\rangle + \langle \delta\hat{\mathbf{S}}_i/S \cdot \delta\hat{\mathbf{S}}_j/S \rangle \; .
\end{equation}
At order $1/S$ one finds that:
\begin{subequations}
    \begin{align}
      \langle \delta \hat{\mathbf{ S}}_i/S \cdot \delta \hat{\mathbf{S}}_{j}/S \rangle &=  \frac{1}{S} \mathcal{R}_{ij}^{\alpha\beta} \langle \delta \hat{\Sigma}^\alpha_{1/2,i} \delta \hat{\Sigma}^\beta_{1/2,i}  \rangle  \\  \mathbf{S}_{cl,i} \cdot \langle \delta \hat{\mathbf{S}}_j / S \rangle &= -\frac{1}{S} \mathcal{R}_{ij}^{zz}  \langle \delta \hat{\Sigma}_{1,j}^z \rangle  \; ,
    \end{align}
\end{subequations}
where $\alpha,\beta = \{ x,y \}$ and $\langle \cdot  \rangle = \bra{0} \cdot \ket{0}$ stands for the Bogoljubov vacuum state average. Therefore, the 2-point correlation function can be written as 
\begin{equation}
\label{eq:corr}
    C_{ij}  = \mathbf{S}_{cl,i}\cdot \mathbf{S}_{cl,j} + \frac{1}{S} c_{ij}^{(1)}
\end{equation}
with    
\begin{equation}
c_{ij}^{(1)} = \mathcal{R}_{ij}^{\alpha\beta} \langle \delta \hat{\Sigma}^\alpha_i \delta \hat{\Sigma}^\beta_i  \rangle - \mathcal{R}_{ij}^{z z} \left( \langle \delta \hat{\Sigma}^z_i \rangle + \langle \delta \hat{\Sigma}^z_j \rangle  \right) \; .
\end{equation}

\begin{figure}
    \centering
    \includegraphics[width=0.6\linewidth]{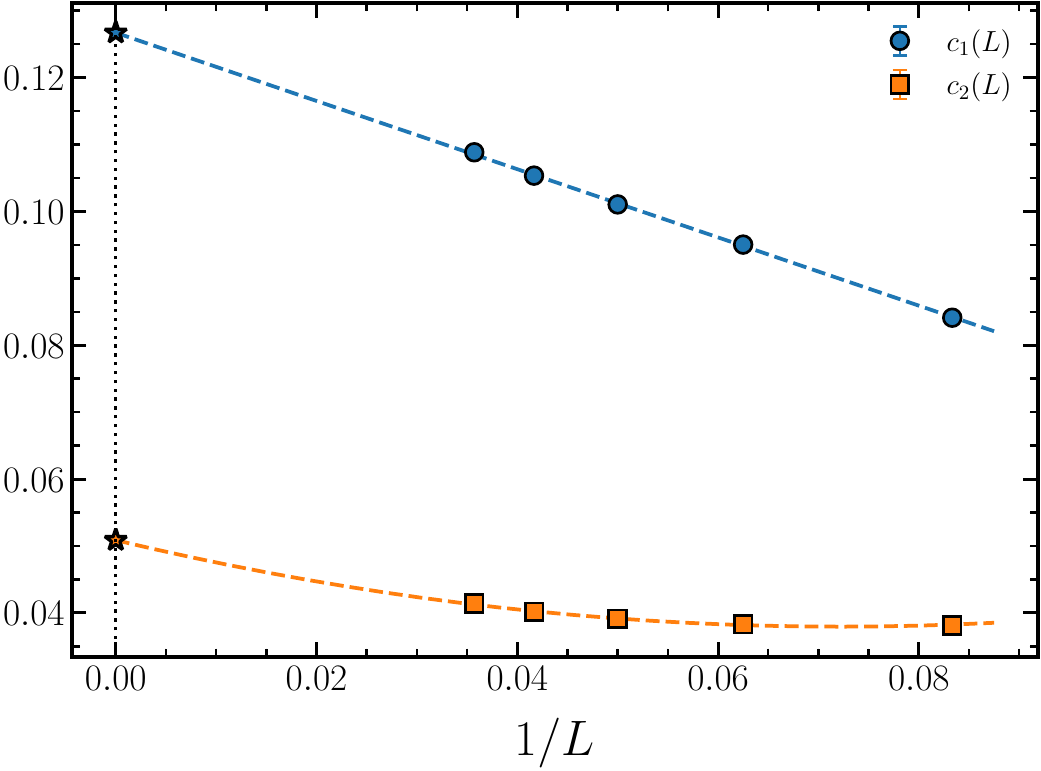}
    \caption{Finite size extrapolations of the corrections to the overlap, for $p=0.5$. Each point is obtained considering an average over approximately $1000$ disorder realizations. The infinite size values are extracted with a quadratic fit in $1/L$ and are respectively $c_1^\infty = 0.127(3)$ and $c_2^\infty = 0.051(1)$.}
    \label{fig:LSWT_extrapolation}
\end{figure}

By squaring the correlation, one finds: 
\begin{equation} \label{eq:corr_sq}
    C_{ij}^2 =  \left(\mathbf{S}_{cl,i}\cdot \mathbf{S}_{cl,j}\right)^2 + \frac{2}{S} \mathbf{S}_{cl,i}\cdot \mathbf{S}_{cl,j}~c_{ij}^{(1)} + \frac{1}{S^2} \left(c_{ij}^{(1)}\right)^2 \; ,
\end{equation}
where we also retain the $1/S^2$ term. It is important to stress that Eq.~\eqref{eq:corr_sq} does not represent the complete result at order $1/S^2$, since magnon interactions also contribute at that order and are not captured by linear spin-wave theory. Rather, it is obtained by computing the correlations within linear spin-wave theory, setting $S=1/2$, and then constructing $Q^2$. The consistency of this procedure is confirmed in the antiferromagnetic limit, where linear spin-wave theory correctly reproduces $Q^2=\mathcal{M}_{\mathrm{N\acute eel}}^4$. We can therefore write:
\begin{equation}
\label{eq:oveSW}
    Q^2(S,L) = \bar{Q}_0^2 - \frac{2}{S} c_1(L) + \frac{1}{S^2} c_2(L),
\end{equation}
where $\bar{Q}_0^2=1/3$ is the classical value,
\begin{subequations}
    \begin{align} \label{eq:oveCorr}
        c_1(L) &= - \frac{1}{L^4} \sum_{ij}^n \mathbf{S}_{cl,i}\cdot \mathbf{S}_{cl,j}~c_{ij}^{(1)} \\
        c_2(L) &= \frac{1}{L^4} \sum_{ij}^n \left(c_{ij}^{(1)}\right)^2 \; .
    \end{align}
\end{subequations}
Notice that, in order to compare the semiclassical prediction for the overlap with the fully quantum one, the right-hand side of Eq.~\eqref{eq:oveSW} has to be multiplied by a factor $S^4$, due to the spin normalization mismatch between the classical case ($|\mathbf{S}_{cl}|^2=1$) and the quantum ($S=1/2$) case. This way, we obtain: $Q^2 = Q_0^2 + \Delta Q^2$, with $Q_0^2=S^4/3$ and 
\begin{equation}
    \Delta Q^2 = - 2 S^3 c_1^\infty + S^2 c_2^\infty \; ,
\end{equation}
where $c_{1,2}^\infty$ denote the extrapolated values of $c_{1,2}(L)$ in the thermodynamic limit (Fig. \ref{fig:LSWT_extrapolation}). 
By taking the square root $Q^2$, one has
\begin{equation}
    Q_{\text{SW}} = \frac{S^2}{\sqrt{3}} \sqrt{1-\frac{6 c_1^\infty}{S} + \frac{3c_2^\infty}{S^2}} \; .
\end{equation}

For completeness, let us also write the explicit expression of $c^{(1)}_{ij}$ in terms of the HP bosonic operators
\begin{equation}
    c_{ij}^{(1)} = \langle \hat a_i^\dagger \hat a_j\rangle \mathcal{A}_{ij} + \langle \hat a_i \hat a_j^\dagger\rangle \mathcal{A}_{ij}^* +  \langle \hat a_i^\dagger \hat a_j^\dagger \rangle \mathcal{B}_{ij} +  \langle \hat a_i \hat a_j\rangle \mathcal{B}_{ij}^* - \mathbf{S}_{cl,i}\cdot \mathbf{S}_{cl,j} (\langle   a_i^\dagger a_i \rangle + \langle   a_j^\dagger a_j \rangle) \; ,
\end{equation}
where 
$\mathcal{R}_{ij}^{z z} = \mathbf{S}_{cl,i}\cdot \mathbf{S}_{cl,j}$. Expectation values of the creation/annihilation operators need to be evaluated on the ground state of the quantum theory. By exploiting the definition of the inverse transformation in Eq.~\eqref{eq:BogandInv} one finds
\begin{align}
     \langle \hat a^\dagger_i \hat a_j \rangle &= (Y^tY^*)_{ij} \qquad \langle \hat a_i \hat a^\dagger_j \rangle =  (X^tX^*)_{ij} \nonumber \\
      \langle \hat a_i \hat a_j \rangle &=  (X^tY^*)_{ij} \qquad
        \langle \hat a^\dagger_i \hat a^\dagger_j \rangle =  (Y^tX^*)_{ij} \; .
\end{align}
These are the only surviving terms, since the only nonzero contributions in the inverse Bogoljubov transformation come from $\langle \hat b \hat b^\dagger \rangle$. The components of $X$ and $Y$ corresponding to the kernel eigenvectors have been excluded from the computation, in line with the discussion at the end of the previous Section.

\bibliography{ref}